\documentstyle[aps,prl,preprint]{revtex}
\newcommand{\be}   {\begin{equation}}
\newcommand{\ee}   {\end{equation}}
\tighten
\input{epsf}
\begin{document}
\draft \title{The potential energy landscape of a model glass former: thermodynamics,
anharmonicities, and finite size effects}
\author{Stephan B\"uchner, Andreas Heuer }
\address{Max-Planck-Institut f\"ur Polymerforschung, Ackermannweg 10, D-55128 Mainz, Germany}
\date{\today} \maketitle
\begin{abstract}

It is possible to formulate the thermodynamics of a glass forming system in
terms of the properties of inherent structures, which correspond to the minima
of the potential energy and build up the potential energy landscape in the
high-dimensional configuration space. In this work we quantitatively apply this
general approach to a simulated model glass-forming system. We systematically
vary the system size between N=20 and N=160. This analysis enables us to
determine for which temperature range the properties of the glass former are
governed by the regions of the configuration space, close to the inherent
structures. Furthermore, we obtain detailed information about the nature of
anharmonic contributions. Moreover, we can explain the presence of finite size
effects in terms of specific properties of the energy landscape. Finally,
determination of the total number of inherent structures for very small systems
enables us to estimate the Kauzmann temperature.

\end{abstract}

\vspace{1cm}
{\bf I. Introduction}
\vspace{1cm}

The physics of glass forming systems is a complex multiparticle problem,  as
reflected, e.g., by the occurrence of non-exponential relaxation or
non-Arrhenius temperature dependence of transport coefficients for most systems
\cite{Richert94,Ediger96}. Beyond phenomenological models like the Gibbs-Adam
model \cite{Adam65} or theoretical approaches like the mode-coupling theory
\cite{Gotze92} computer simulations have become increasingly important to yield
additional insight into the nature of the glass transition from a microscopic
viewpoint.

A fruitful approach is  the concept of the potential energy landscape (PEL)
\cite{Goldstein69,Angell95,Stillinger95}. In this approach the total system is
regarded as a single point moving in the high-dimensional configuration space on
a time-independent landscape, representing the potential energy. To a large
extent the topography of the PEL is characterized by the local energy minima,
also denoted {\it inherent structures}. Although the analysis of inherent
structures has been applied to several problems \cite{Stillinger83,Ohmine88,Jonsson88,Sciortino91,Tanaka96}. Until now, only limited quantitative
information is available concerning the PEL of glass forming systems. This is at
least partly related to the fact that the number of inherent structures
exponentially increases with system size so that a complete enumeration is only
possible for very small systems. This has been demonstrated for small clusters
\cite{Berry93,Miller99} as well as for monatomic Lennard-Jones systems with
periodic boundary conditions for up to 32 particles \cite{Heuer97f,Angelani98}.
Since monatomic systems tend to crystallize even on computer time scales it has
become common to use binary rather than monatomic systems to suppress
crystallization \cite{Stillinger85,Heuer93,Kob95}. For these systems as well as
for slightly larger monatomic systems, however, a complete enumeration is no
longer possible so that one has to resort to an appropriate statistical
analysis. Such an approach has been used in \cite{Ball96} where the distribution
of local minima for a KCl cluster is determined.

A major question, which has become of increasing importance, is the relevance of
the PEL \cite{Sastry98,Schroder99,Buechner99}. In a trivial sense the PEL just
reflects the full potential energy of the system and is therefore always
relevant. In a less trivial sense one may ask {\it whether the physics of the
system is governed only by the part of the configuration space close to the
inherent structures}. In a recent work it has been shown for a Lennard Jones
system that exactly for the temperature region $T < T_r$, for which typical
features like the non-exponentiality of the structural relaxation are observed
also the average energy of inherent structures depends on temperature
\cite{Sastry98}. From this observation the authors concluded that the PEL is
indeed relevant for temperatures below some temperature $T_r$. Interestingly,
$T_r$ is significantly larger than the critical temperature $T_c$ of the
mode-coupling theory \cite{Gotze92}. In Ref. \cite{Schroder99} it was shown that
close to $T_c$ the dynamics of the model glass former can be basically viewed as
a superposition of hopping processes between the different inherent structures
and harmonic vibrations around them. This is a very direct piece of evidence for
the relevance of the PEL in the sense mentioned above. Furthermore it could be
shown explicitly that the presence of fast and slow regions in a glass former,
and thus the presence of non-exponential relaxation, can be attributed to the
topography of the PEL \cite{Buechner99}. Also the relevance of the PEL for aging
has been recently demonstrated \cite{Kob99}.

If the system mainly resides close to the inherent structures of the PEL, the
potential energy can be described in harmonic approximation around these
inherent structures, respectively. Therefore our question concerning the
relevance of the PEL can be reformulated by asking to which degree the
properties of the system can be described in harmonic approximation.  If the
system always resides in a single minimum the degree of anharmonicity can be
simply determined, e.g. by analysis of the temperature dependence of the mean
fluctuations around an inherent structure \cite{Sastry98}. At higher
temperatures for which the residence time close to a single inherent structure
may be small these approaches become unreliable.

In this paper we want to show that computer simulations can be used to yield a
variety of information about the PEL.  The main ingredients of our simulations
have been already proposed by Stillinger and coworkers
\cite{Stillinger83,Stillinger88,Stillinger95}. First, we use their algorithm,
combining standard molecular dynamics (MD) simulation with regular quenching of
the potential energy. Second, we adapt their formulation of the partition
function of the total system in terms of the properties of the individual
inherent structures.  Combination of both ingredients will yield quantitative
information about the partition function and thus about the thermodynamics of
the system. More specifically the following aspects will be analysed: (i)
Characterization of the PEL in terms of the density of inherent structures (ii)
Dependence of the PEL on system size and comparison with scaling relations one
would expect for sufficiently large systems. (iii) Quantification of anharmonic
contributions. (iv) Connection of the PEL to dynamic properties. (v)
Consequences for thermodynamic properties like the specific heat and the
presence of a Kauzmann temperature. In the field of clusters similar approaches
have been already applied \cite{Rose93,Doye95}.

The organization of this paper is as follows. In Sect. II we present a detailed
outline of the conceptual background of the approach chosen in this work. Sect.
III contains a description of our simulation method and  the model system. In
Sect. IV the dynamics and the structure is characterized via standard Molecular
Dynamics (MD) simulations. In Sect. V we present the main results of our
simulations with respect to properties of the PEL. The discussion of the
implications of these results can be found in Sect. VI.

\vspace{1cm}
{\bf II. Partition function of glass forming systems}
\vspace{1cm}

In this Section we present the conceptual background applied in this work and
introduce the notations used thereafter. This outline is rather detailed in
order to make the implications of this approach as clear as possible. Starting
from the distribution function of potential energies ${\cal G}(E)$,
characterizing the total configuration space, the configurational contribution
of the canonical partition function $Z(T)$ can be expressed as
\be
\label{z_tradition}
Z(T) = \int_{-\infty}^{\infty} dE {\cal G}(E) \exp(-\beta E)
\ee
where $\beta = 1/T (k_B \equiv 1)$. No specific information about inherent
structures is contained. In case that the physics is mainly determined by the
inherent structures and their close neighborhood, respectively, it may be more
informative to express the partition function in terms of the properties of the
inherent structures. The main idea is to split the total configuration space in
contributions corresponding to the different inherent structures $i$ with energy
$\epsilon_i$, i.e. the  minima of the potential energy of the system. Each
inherent structure is surrounded by a so-called basin of attraction $\Omega_i$.
It is defined as the set of all configurations which end up as the inherent
structure $i$ upon energy minimization. Since the mapping of configurations on
inherent structures via enery minimization is unique (except for a set of
configurations with measure zero, corresponding to the saddle points of the PEL)
the total configuration space can be decomposed in disjoint partitions
$\Omega_i$. Then   $Z(T)$ can
 be written as the sum over the individual partition functions
$Z_i(T)$, i.e. $Z(T) = \sum Z_i(T)$, where the $Z_i(T)$ are defined as
\be
Z_i(T) \equiv \int_{\Omega_i} d\vec{r_1}...d\vec{r_N}\exp(-\beta
V(\vec{r_1},...,\vec{r_N})).
\ee
The $\{\vec{r}_j\}$ denote the positions of the $N$ particles of the system and
the integration is over the basin of attraction of the i-th inherent structure.

For the final calculation of the partition function it is helpful to rewrite
the summation over all inherent structures by combining all contributions of
inherent structures with the same energy $\epsilon$. For this purpose we
introduce the partition function $Z(\epsilon,T)$, defined as
\be
Z(\epsilon,T) = \sum_i Z_i(T) \delta(\epsilon - \epsilon_i),
\ee
such that
\be
Z(T) = \int d\epsilon Z(\epsilon,T).
\ee

On a qualitative level $Z(\epsilon,T)$ is a measure for the probability that a
configuration at temperature $T$ belongs to a basin of attraction of an inherent
structure with energy $\epsilon$. Actually, as discussed in the next Section, it
is this quantity $Z(\epsilon,T)$ which, apart from a proportionality factor,  we
can extract from our simulations. If $G(\epsilon)$ denotes the number of
inherent structures with energy $\epsilon$ we can furthermore introduce the
average value   $z(\epsilon,T)$ for all inherent structures with energy
$\epsilon$ via
\be
\label{z_intro}
z(\epsilon,T)  \equiv Z(\epsilon,T)/ G(\epsilon).
\ee

 In general, $Z(\epsilon,T)$ may be a very
complicated function of $T$ and $\epsilon$. In the limit of low
temperatures, however, it is reasonable to assume that apart from
the energy $\epsilon_i$ itself the individual partition functions
$Z_i$ are mainly determined by the harmonic contributions, i.e.
$Z_i(T)
\approx  \exp(-\beta \epsilon_i)  Z_i^{harm}(T)$, so that  in general it is
helpful to take into account harmonic and anharmonic contributions individually.
The harmonic contributions are given by
\be
Z_i^{harm}(T) \equiv
\prod_j
\left ( \frac{2\pi T}{\nu_{j,i}}
 \right )^{1/2} \equiv Y_i^{harm} T^{(3N-3)/2}
\ee
where $\nu_{j,i}$ denote the $3N-3$ positive eigenvalues of the force matrix
evaluated for the i-th inherent structure. Note that the temperature dependence
of the vibrational partition function $Z_i^{harm}(T)$ is simply given by the
factor $T^{(3N-3)/2}$ whereas $Y_i^{harm}$ contains the temperature-independent
information about the harmonic modes around this inherent structure. In analogy
to above we define $y^{harm}(\epsilon)$ as the average of the $Y_i^{harm}$ over
all inherent structures  with energy $\epsilon$. Then we can write
\be
\label{anh}
z(\epsilon,T) \equiv \exp(-\beta \epsilon) y^{harm}(\epsilon)  T^{(3N-3)/2}
z^{anh}(\epsilon,T),
\ee
 thus introducing  the term $z^{anh}(\epsilon,T)$, accounting for the
anharmonic corrections. By definition one has $z^{anh}(\epsilon,T)=1$ for
sufficiently low temperatures. In literatur, phenomenological expressions for
the description of anharmonic contributions can be found; see, e.g.,
\cite{Rose93,Doye95}. Finally, the total partition function can be expressed as
\be
\label{z_new}
Z(T) = T^{(3N-3)/2} \int d\epsilon G(\epsilon) y^{harm}(\epsilon)
z^{anh}(\epsilon,T)\exp(-\beta \epsilon)
\ee

Since all thermodynamic quantities can be derived from knowledge of the
partition function it is evident from Eq.\ref{z_new} that it is not the density
of inherent structures $G(\epsilon)$ alone which determines the properties of
the system. At sufficiently low temperatures it is rather the product
$y^{harm}(\epsilon) G(\epsilon)$ which is relevant. We denote this product {\it
effective density} $G_{eff}(\epsilon)$, i.e.
\be
\label{geff_def}
G_{eff}(\epsilon) \equiv y^{harm}(\epsilon) G(\epsilon).
\ee
It can be determined from $Z(\epsilon,T)$ via
\be
\label{geff}
G_{eff}(\epsilon) = T^{-(3N-3)/2} Z(\epsilon,T) \exp(\beta
\epsilon)/z^{anh}(\epsilon,T).
\ee
Thus for sufficiently low temperatures for which $ z^{anh}(\epsilon,T)=1$ we can
directly obtain the effective density of states from a reweighting of the $
Z(\epsilon,T)$ with the inverse Boltzmann factor. The resulting effective
density $G_{eff}(\epsilon)$ is independent of temperature. In practice one has
to determine $Z(\epsilon,T)$ for several temperatures in order to obtain
$G_{eff}(\epsilon)$ for a wide range of energies.

Finally, the total partition function can be expressed in terms of the effective
density via
\be
\label{partition}
Z(T) = T^{(3N-3)/2} \int d\epsilon G_{eff}(\epsilon) z^{anh}(\epsilon,T)
 \exp(-\beta \epsilon).
\ee
Despite the formal similarity with Eq.\ref{z_tradition} the present approach is
based on a description in terms of the distribution of inherent structures in
contrast to an overall description of the PEL, expressed in
Eq.\ref{z_tradition}. The main advantage of the present approach is the
possibility to uniquely identify anharmonic contributions. A straightforward way
to do this is to calculate a thermodynamic quantity like the specific heat, on
the one hand, directly from the MD configurations and, on the other hand, from
Eqs. \ref{geff} and \ref{partition} with $z^{anh}(\epsilon,T)=1$, i.e. using the
harmonic approximation. Deviations between both approaches can be uniquely
attributed to anharmonic contributions, i.e. invalidation of the relation
$z^{anh}(\epsilon,T)=1$.

Finally we would like to mention that there exist alternative approaches to
formulate the thermodynamics via a combination of constant energy MD simulations
and quenching from which the energy density ${\cal G}(E)$ for different systems
has been estimated; see, e.g., Ref. \cite{Doye95}.

\vspace{1cm}
{\bf  III. Methods}
\vspace{1cm}

We studied a binary Lennard-Jones (LJ)-type system. The mutual interactions  are
chosen such that the interaction between unlike particles is favoured, thus
avoiding crystallisation for an appropriately chosen mixing ratio. The pairwise
interaction potential has been proposed by Stillinger and Weber
\cite{Stillinger85}
\be
V_{ij}(r_{ij}) = C \epsilon_{\kappa(i)\kappa(j)}
[(r_{ij}/\sigma_{\kappa(i)\kappa(j)})^{-12}-1]\exp[(r_{ij}/\sigma_{\kappa(i)\kappa(j)}-a)^{-1}];
r_{ij} < \sigma_{\kappa(i)\kappa(j)}
\ee
and zero otherwise.  Here $\kappa(i) \in \{A,B\}$ indicates whether the i-th
particle is an A or a B type particle. The parameters are $C = 8.805977,
a=1.652194, \epsilon_{AA}=1,
\sigma_{AA}=1.0,\epsilon_{AB}=1.5 \epsilon_{AA}, \sigma_{AB} = 2.00/2.49
\sigma_{AA}, \epsilon_{BB}=0.5 \epsilon_{AA}, \sigma_{AB} = 2.20/2.49
\sigma_{AA}.$ The system contains 80\% A-particles and 20\% B-particles.
Energy and length units are given in units of $\epsilon_{AA}$ and $\sigma_{AA}$.
Finally, the time unit is $\sqrt{m_A
\sigma_{AA}^2/\epsilon_{AA}}$. As
compared to a LJ potential with a standard cutoff at $r
= 2.5$ (in LJ units) this potential is more short-ranged. We performed
simulations at constant density $\rho = 1.204$, temperatures ranging from 0.667
to 2.5, and system sizes between $N=20$ and $N=160$. The glass former was
propagated at a given temperature $T$ via standard molecular dynamics (MD)
techniques, using the velocity form of the Verlet algorithm with  time steps
depending on temperature but smaller than 0.00125. The temperature was kept
constant via velocity rescaling, i.e. by using a constant kinetic energy during
our simulation run. Alternatively, we applied the Nose equations of motion
\cite{Nose83}, with no significant variations for the quantities discussed in
this work. We checked that upon shifting the temperature scale by 30\% to lower
temperatures the present Lennard-Jones type model can be mapped to the model
presented in \cite{Kob95} for temperatures in the supercooled regime.

First we performed standard MD simulations at different temperatures yielding
information about the relaxation properties like the structural ($\alpha$)
relaxation time. To obtain information about the PEL we calculated inherent
structures by the conjugate gradient minimization technique. The procedure was
such that during an MD run at constant temperature the system was regularly
minimized and after each minimization procedure the MD run was continued with
the same configuration and momenta as before the minimization. This is
schematically shown in Fig.1. The thick line corresponds to the MD trajectory,
the thin lines sketch the path the system takes upon quenching. During every
minimization process the MD configuration is mapped on the inherent structure,
whose basin of attraction comprises the MD configuration. On average we
performed 20 minimization procedures during one $\alpha$ relaxation time.

The probability that an arbitrary MD configuration belongs to a
basin of attraction of the i-th inherent structure  is given by
$Z_i(T)/Z(T)$. Therefore the probability $P(\epsilon,T)$ to find an
inherent structure with energy $\epsilon$ (at constant temperature)
by the above procedure is given by $Z(\epsilon,T)/Z(T)$. This is
the key feature which according to the outline of Sect. II allows
us to extract thermodynamic properties from this type of procedure.

\vspace{1cm}
{\bf IV. Dynamics and Structure}
\vspace{1cm}

In this Section we present results, characterizing the dynamics of
our LJ-type system for different system sizes and different
temperatures. The dynamics can be conveniently described by the
intermediate incoherent scattering function $S(q,t)$ which is
defined as
\be
S(q,t) = \frac{1}{N} \sum_i \cos(\vec{q} (\vec{r}_i(t) -
\vec{r}_i(0))
\ee
where $\vec{q}$ denotes the scattering vector and $\vec{r}(t)
- \vec{r}(0) $ the displacement of a particle during time $t$.
Here we restrict ourselves to the A particles. For isotropic
systems only the absolute value $q$ of the scattering vector is
relevant. In what follows we take a value of $q$ close to the first
maximum of the structure factor, i.e. the inverse typical particle
distance ($q=7.251$). In Fig.2 we show $S(q,t)$ for $T=0.66$ for
different system sizes $N$. For all sizes one can clearly see the
two-stage relaxation ( fast $\beta$ and $\alpha$ process) as
predicted by the mode-coupling theory. Starting from large values
of $N$ only minor variations of $S(q,t)$ occur for $N
\ge 60$. The most significant observation is that strong finite
size effects occur for $N < 60$. In this regime the relaxation time
strongly increases with decreasing system size. However, even for
$N=20$ one observes on a qualitative level, the same two-step
relaxation process as for large system sizes. We checked  for
$T=0.883$ that also for system sizes between $N=160$ and $N=480$ no
systematic variation with $N$ is observed.

In Fig.3 we show the temperature dependence of $S(q,t)$ for $N=60$. As already
known from many different experiments and simulations the $\alpha$-relaxation
time strongly increases with decreasing temperature. In Fig.4 we display the
$\alpha$-relaxation time for a large part of the $(T,N)$ plane. It is defined
via $S(q,\tau_\alpha) = 1/e$. One can clearly see that for all temperatures
analysed in this work strong finite size effects start to play a role for $N <
60$. The apparent step in relaxation times between $N=60$ and $N=40$ decreases
with increasing temperature. Interestingly, for $N=20$ as well as for $N=40$ one
observes an Arrhenius temperature dependence for low temperatures. In contrast,
for large $N$ one observes a continuously increasing apparent activation energy,
in agreement with typical experimental observations on  fragile glass formers.
It has been already reported earlier for a monatomic Lennard-Jones-type system
with 32 particles that at low temperatures the relaxation has an Arrhenius
temperature dependence \cite{Heuer97f}. For that system the low-temperature
activation energy could be related to an effective barrier of the PEL around a
particular inherent structure with a low energy which was visited very often at
low temperatures. A similar reason will be discussed below for the present case.

As demonstrated in Fig.5, also the pair correlation function $g(r)$ between
particles of the minority component B indicates significant finite size effects
at the lowest temperature.  Again, only for $N
\ge 60$ the bulk limit is approximately  reached. This indicates that there is a
common reason for finite size effects, relevant for static and dynamic
properties. In contrast, only very mild finite size effects can be observed
between particles of the majority component A.

\vspace{1cm}
{\bf V. The potential energy landscape}
\vspace{1cm}

Based on the algorithm discussed in Sect.III we analysed runs with lengths
between 300 and 1000 $t_\alpha$. For system size $N=60$ and for three
representative temperatures ($T=1.667, 0.833, 0.667$)  we show
$\epsilon(t)$-curves in Fig.6, reflecting the energy variation of the inherent
structures with time. Closer inspection of the $\epsilon(t)$ time series for
$T=0.833$ and $T=0.667$ reveals that there are long periods of time during which
the system is jumping back and forth between a small number of inherent
structures. This scenario can be interpreted in terms of valleys on the PEL in
which the system is caught for some time \cite{Buechner99}. Here we concentrate
on the statistics of the inherent structures.

In Fig.7 we plot the average value of the energy of inherent
structures, denoted $\langle
\epsilon
\rangle_T$, for different temperatures. This plot is similar to
the curves shown in Ref.\cite{Sastry98}. The temperature variation for $T=0.833,
0.714, 0.667$ is consistent with a $1/T$ behavior whereas at high temperatures
the temperature dependence becomes weaker. In Ref. \cite{Sastry98} the authors
additionally observed a low-temperature plateau, which, however, was exclusively
related to non-equilibrium effects and correspondingly strongly depends on the
thermal history. Here, we restrict ourselves to the regime of equilibrium
dynamics. In order to get a closer understanding of this temperature dependence
we have determined not only the average value but also the whole probability
curve $P(\epsilon,T)$ that at temperature $T$ one observes an inherent structure
with energy $\epsilon$. As shown in Fig.8 the distribution $P(\epsilon,T)$
continuously shifts to lower energies when decreasing the temperature but does
not change its shape or width. Our goal is to derive the effective density
$G_{eff}(\epsilon)$, see Eq.\ref{geff_def}, of inherent structures from
knowledge of $P(\epsilon,T)$. Since $Z(\epsilon,T)$ is proportional to
$P(\epsilon,T)$, the effective density $G_{eff}(\epsilon)$ can in principle be
determined from Eq.\ref{geff} except for a proportionality constant which only
depends on temperature, i.e.
\be
\label{f(T)}
G_{eff}(\epsilon) z^{anh}(\epsilon,T) \propto P(\epsilon,T)
\exp(\beta \epsilon).
\ee
Obviously, application of  Eq.\ref{f(T)} requires  knowledge of
$z^{anh}(\epsilon,T)$, which in general is not available. If, however,
$z^{anh}(\epsilon,T)$ does not depend on $\epsilon$ (which trivially holds in
the low temperature limit where $z^{anh} \equiv 1$ but, of course, is a more
general condition) it can be included in the proportionality constant. Then the
$\epsilon$-dependence of $G_{eff}(\epsilon)$ can be determined from
multiplication of $P(\epsilon,T)$ with an inverse Boltzmann factor except for a
proportionality constant. In principle a single temperature is sufficient to
obtain $G_{eff}(\epsilon)$. However, as already shown in Fig.8, for different
temperatures $P(\epsilon,T)$ is distributed around different energies. Therefore
in practice it is necessary to combine the simulations at different temperatures
to obtain larger parts of the $G_{eff}(\epsilon)$ distribution; see e.g.
\cite{Labastie90}. The relative proportionality factors are determined by the
condition that $G_{eff}(\epsilon)$, extracted from different temperatures,
should be identical in the overlap region. This procedure is performed in Fig.9.
Obviously, for the three lower temperatures $T=0.667;0.714;0.833$ the overlap is
close to be perfect. It remains a single unknown proportionality constant which
we accounted for by plotting $G_{eff}(\epsilon)/G_{eff}(\epsilon_{0})$  where
$\epsilon_{0}$ is the lowest energy found during the simulations. Interestingly,
the $G_{eff}(\epsilon)$ curves, obtained from the high-temperature simulations
($T=1.667$ and $T=2.5$) do not overlap with the low-temperature data. As
discussed above this directly indicates that at high temperatures anharmonic
contributions are present and furthermore depend, as expressed by
$z^{anh}(\epsilon,T)$,  on energy $\epsilon$. The $G_{eff}(\epsilon)$ curves
were shifted such that they agree with the low-temperature curves in the region
of large $\epsilon$. No mapping was possible for the low $\epsilon$ region. This
behavior as well as the consequences will be discussed in Sect. VI.

The energy dependence of the effective density can be excellently fitted by a
gaussian distribution $\exp(-(\epsilon - {\epsilon_{max}})^2/2\sigma^2)$ with
${\epsilon_{max}} = -5.6 N$ and $\sigma^2 = 0.3 N$. A gaussian distribution
naturally occurs in the limit of very large $N$. In this limit it is reasonable
to assume that the total system can be decomposed into only weakly interacting
subsystems so that the total energy is a sum of weakly correlated energy
contributions. According to the central limiting theorem this naturally results
in a gaussian distribution. It is nevertheless surprising that already for
$N=60$ the gaussian distribution is  a very good approximation to the true
distribution although such a small system definitely cannot be decomposed into
only weakly interacting subsystems. As shown below even for $N=20$ one obtains a
distribution function which closely resembles a gaussian distribution.

Based on the knowledge of $G_{eff}(\epsilon)$ it is possible to
estimate $\langle \epsilon \rangle_T$ in harmonic approximation;
see Sect.II. This results in
\be
\langle \epsilon \rangle_T^{harm} = \epsilon_{max} - \frac{\sigma^2}{T}.
\ee
The resulting curve for $N=60$ is also included in Fig.7. Whereas in the
low-temperature regime one has $\langle \epsilon \rangle_T^{harm} \approx
\langle \epsilon
\rangle_T$, both curves deviate at high temperatures. As discussed in Sect.VI this is
a direct consequence of the apparent temperature dependence of
$G_{eff}(\epsilon)$ for high temperatures, discussed above.

We also checked the $\epsilon$-dependence of $y^{harm}(\epsilon)$. This is
essential in order to estimate the density of inherent states $G(\epsilon)$ from
$G_{eff}(\epsilon)$. Again this analysis can be performed for different
temperatures. To be specific, we calculated the average value of
$\ln(Y_i^{harm})$ for all inherent structures with energy $\epsilon_i =
\epsilon$, obtained from our quenching procedure. Formally, the resulting
expectation value can be written as
\be
\langle \ln (y^{harm}) \rangle (\epsilon) =
\frac{\sum_i \delta (\epsilon - \epsilon_i) \ln(y_i^{harm})y_i^{harm} z^{anh}_i(T)}
{\sum_i \delta (\epsilon - \epsilon_i) y_i^{harm} z^{anh}_i(T)}
\ee
For low temperatures where anharmonic effects can be neglected one expects
temperature independent expectation values $\langle \ln (y^{harm}) \rangle
(\epsilon)$. The results are shown in Fig.10. For the three lower temperatures
no significant temperature dependence can be observed. Interestingly, a weak
dependence on $\epsilon$ is observed: higher energies correspond to smaller
values of  $\langle
\ln y^{harm}\rangle$ and thus to larger harmonic force constants.
This result is consistent with recent simulations on small monatomic LJ-systems
\cite{Angelani99}. Due to the $\epsilon$-dependence of $y^{harm}(\epsilon)$ the
density $G(\epsilon)$ and the effective density $G_{eff}(\epsilon)$ slightly
deviate from each other. It turns out, however, that the variances of
$G(\epsilon)$ and $G_{eff}(\epsilon)$ differ by less than $10\%$. In what
follows this effect is neglected and we choose $G(\epsilon)
\propto G_{eff}(\epsilon)$. Interestingly, the values of $y^{harm}(\epsilon)$ are shifted
to smaller values if the inherent structures are analysed obtained from the high
temperature simulations ($T=1.667$ and $T = 2.5$). Again, this is a clear
signature of anharmonic effects. Thus the temperature dependence of the average
harmonic partition function (see \cite{Kob99}), averaged over all inherent
structures at a given temperature, has two contributions, (i) the
$\epsilon$-dependence which via the temperature dependence of the average energy
of inherent structures $\langle
\epsilon \rangle_T$ translates into a temperature dependence of the harmonic partition function
 and (ii) the temperature dependent anharmonic effects.

In order to independently check the  degree of gaussianity of
$G_{eff}(\epsilon)$ one may check the temperature dependence of the energy
variance $\sigma_P^2(T)$ of $P(\epsilon ,T)$. In case of a gaussian distribution
one expects $\sigma_P^2(T)
=
\sigma^2$. In Fig.11 we display $\sigma_P^2(T)/N$. Extending the results, reported above, we have also
included the data for different system sizes $N$. We first concentrate on the
data for $N=60$ and for reasons, mentioned above, concentrate on the three
low-temperature data. It turns out that the energy variance is indeed constant,
and is consistent with the value, directly obtained from $G_{eff}(\epsilon)$. It
is very illuminating to discuss the N-dependence of $\sigma$. In the macroscopic
limit $N \rightarrow
\infty$ application of the central limiting theorem suggests
${\epsilon}_{max} \propto N$ and $\sigma^2 \propto N$. Interestingly, within
statistical error all data for $\sigma^2/N$ agree for $N
\ge 60$.

Systems with size smaller than $N=60$ display significant finite size effects in
terms of the distribution of inherent structures. Interestingly, the variance
decreases with decreasing temperature for $N=20$ and $N=40$. The reason for this
temperature dependence can be directly understood from the plot of
$\epsilon(t)/N$ for $N=20$ at $T=0.667$; see Fig.12a. It becomes evident that it
is a single inherent structure which dominates the distribution of inherent
structures. This dominance directly explains the decreasing variance. The
frequent occurrence of this low-energy inherent structure  does not mean that
the system does no longer relax. In order to clarify this point we introduce the
mobility $\mu(t)$ via
\be
\mu(t) =
\sum_{i=1}^N(\vec{r}_i(t+t_\alpha/2) - \vec{r}_i(t+t_\alpha/2))^2.
\ee
It denotes the mobility at time $t$ on the time-scale  of the
$\alpha$-relaxation time $t_\alpha$. As shown in Fig.12b there exists times when
the system is very mobile. Indeed, at these times the system leaves its
ground-state type structure and after larger rearrangements ends up in a new
configuration which except for permutations and some translational shift is
identical to the former structure. During the other times the system only jumps
between a small number of inherent structures, resulting in a small value of the
mobility $\mu(t)$. For comparison we also show the time dependence of the true
potential energy $E(t)$ for the same run, directly obtained for the MD
trajectory; see Fig.12c. Here, no specific features can be observed. This
examplifies the large information content when analyzing inherent structures
rather than the original MD configurations.

The observation that the low-temperature dynamics of the $N=20$ sample is
dominated by a single inherent structure gives a straightforward interpretation
of the dependence of the pair correlation function on $N$ since the structure of
$g_{BB}(r)$ is also dominated by this inherent structure. Calculating
$g_{BB}(r)$ for the corresponding inherent structure, shown in Fig.13, reveals
that there only exists a single distance between the four B-particles. This type
of behavior can be understood from the Hamiltonian of the system. Since A-A and
A-B contacts are preferred due to the large binding energy the system tries to
maximize the distance between B particles. Indeed, the distance between B
particles is much larger than the optimum binding distance between B particles.
For $N=60$ all distinct features have disappeared.

In a next step we want to analyze the dependence of $G_{eff}(\epsilon)$ on
system size and particle composition. It has been argued in literature that for
large $N$ the number of inherent structures should scale like $\exp(\alpha N)$
where the constant $\alpha$ depends on the type of system. Of course, for small
$N$ the value of $\alpha$ may depend on $N$.  Since to a very good approximation
$G_{eff}(\epsilon) \propto G(\epsilon)$ (see above) also the latter distribution
can be described as a gaussian. For small systems where we can identify an
inherent structure  with minimum energy $\epsilon_{min}$, determination of the
absolute value of the number of inherent structures is possible. Here this is
the case for $N=20$ and $N=30$. Some technical points enter a quantitative
analysis. We have introduced $G(\epsilon)$ as the {\it density} of inherent
structures such that $G(\epsilon) d\epsilon$ denotes the number of inherent
structures in the interval $[\epsilon-d\epsilon/2,\epsilon+d\epsilon/2]$. The
normalization is achieved by setting $G(\epsilon_{min}) = 1$. Since we are
dealing with binary systems we can to a very good approximation neglect any
contributions which arise due to intrinsic symmetries of the configurations. A
similar analysis has been performed in Ref.\cite{Rose93} for the case of
(KCl)$_{32}$. In that work two gaussians  rather than a single gaussian were
needed to fit $P(\epsilon,T)$ and thus $G(\epsilon)$.

For both values of $N$ the resulting $G(\epsilon)$ curves are plotted in Fig.14.
On a qualitative level one can already see that the number of inherent
structures is by orders of magnitudes larger for $N=30$ than for $N=20$. For a
quantitative analysis of the number of inherent structures we assume that the
description of $G(\epsilon)$ as a gaussian also holds for $\epsilon >
\epsilon_{max}$. From the present simulations these inherent structures are not accessible
because they are unfavoured from the entropic as well as from the energetic
point of view. In a previous work, however, it has been shown for a monatomic
Lennard-Jones-type system with 32 particle that the distribution of inherent
structures(for that system approximately 400 inherent structures were found) can
indeed be qualitatively described by a gaussian also for the high-energy wing
\cite{Heuer97f}. For a gaussian the number of inherent structures $N_{is}$ are
related to $G(\epsilon)$ via
\be
N_{is} = G({\epsilon}_{max}) \sqrt{2\pi\sigma^2}.
\ee
From this relation we can estimate $\alpha(N=20)= 0.53 \pm 0.02 $ and
$\alpha(N=30) = 0.70 \pm 0.05$. Thus the value of $\alpha$ slightly increases
when going from $N=20$ to $N=30$. Unfortunately, this value cannot be estimated
for larger $N$ by the present approach since no information about the inherent
structure with the lowest energy  is available so that normalization of
$G(\epsilon)$ is not possible.

In Fig.15 we show $G(\epsilon)$ for two different compositions ($N_A=25,N_B=5$
vs. $N_A=24,N_B=6$). Starting from a monatomic system and having only slightly
different properties of A as compared to B particles one expects that the number
of inherent structures with different energies is proportional to the binomial
coefficient $N!/N_A!N_B!$. According to this argument one would expect that for
the standard composition ($N_A=24,N_B=6$) the number of inherent structures is
approximately $25/6\approx 4$ times higher. Determination of $\alpha$ yields
$\alpha(24:6) = 0.70 \pm 0.05$ and $\alpha(25:5) = 0.58 \pm 0.04$. The number of
inherent structures has therefore increased by a factor of approximately
$\exp(\Delta
\alpha N) =
\exp(0.12
\times 30) \approx 36$. Thus the
increase of the number of inherent structures is larger than a factor of four,
following from purely statistical considerations. Having in mind that this
argument only holds for nearly identical A and B particles, the present case of
two significantly different species may be a source for additional disorder and
thus for an increased number of inherent structures \cite{Stillinger99}.

Finally we calculate the specific heat. From the partition function in
Eq.\ref{partition} one can calculate the specific heat $c(T)$ per particle in
harmonic approximation
\be
\label{charm}
c^{harm}(T) = 3 + \sigma^2/(NT^2).
\ee
The second term expresses the configurational contributions. In Fig.16 this is
compared with the specific heat,  obtained from our simulations via the
fluctuations of the potential energy, i.e.
\be
c(T) = 3/2 + \frac{\langle (E - \langle E \rangle )^2 \rangle }{NT^2}.
\ee
We have plotted the average specific heat for $N=60,80,120,160$, which within
statistical error are identical. It turns out that the agreement between both
curves is good for the three lower temperatures. Interestingly, the simulated
data are significantly larger than $c^{harm}$, indicating the relevance of
anharmonic terms. In contrast, for the higher temperatures $T=1.667$ and $T=2.5$
the specific heat is much smaller than $c^{harm}(T)$. For $T \rightarrow \infty$
the specific heat will approach the ideal gas limit 3/2.

\vspace{1cm}
{\bf VI. Discussion}
\vspace{1cm}

{\bf Anharmonicity}

For several observables discussed above predictions can be made in harmonic
approximation which are based on the effective density $G_{eff}(\epsilon)$,
determined at sufficiently low temperatures on the basis of $P(\epsilon,T)$.
Thus any deviations from this prediction can be directly related to anharmonic
contributions.  In this Section we try to characterize the anharmonic
contributions. Specifically we observe anharmonic contributions for the
following observables: (i) For the two highest temperatures it was not possible
to determine $G_{eff}(\epsilon)$ on the basis of $P(\epsilon,T)$; see Fig.9.
Qualitatively the plot in Fig.9 indicates that at high temperatures the
low-energy inherent structures are found more often than expected from
extrapolation of the low-temperature data. It will be discussed below why
anharmonic effects may lead to this effect. In contrast, for the three lower
temperatures scaling was possible, thus enabling us to determine the effective
density $G_{eff}(\epsilon)$.  From the observed $G_{eff}(\epsilon)$, which
closely resembles a gaussian distribution, one expects a linear increase of
$\langle
\epsilon\rangle_T$ with inverse temperature as long as anharmonic effects are negligible.
However, since due to anharmonic effects low-energy inherent structures were
found too often at high temperatures the average energy of inherent structures
$\langle \epsilon\rangle_T$ must be smaller than expected. In agreement with the
results of Sastry et al. we indeed observe a much weaker increase of $\langle
\epsilon\rangle_T$ for the two highest temperatures; see Fig.7. Thus it is the effect of
anharmonicities which dominates the temperature behavior of $\langle
\epsilon\rangle_T$ at high temperatures. Note that this type of conclusion can be drawn since
we have measured the total distribution function $P(\epsilon,T)$ rather than
only its first moment. (ii) For all temperatures there were small but
significant deviations of the specific heat. Whereas for the three lower
temperatures the anharmonicities give rise to a slightly increased specific
heat, for the higher temperatures one observes a dramatic decrease. (iii) The
expectation values $\langle \ln y^{harm} \rangle (\epsilon)$ depend on
temperature which again can be only rationalized by anharmonic effects.

These effects of anharmonicity, found in our simulations, can be rationalized on
the basis of a simple model potential
\be
V(x) = (1/2)ax^2 - (1/4)b_1 x^4 - (1/6) b_2 x^6
\ee
with the minimum at $x=0$ ($a,b_1,b_2 > 0)$ and maxima at $\pm x_c$ so that its
basin of attraction is the interval $[-x_c,x_c]$.  It is sketched in Fig.17. For
reasons of simplicity we restrict ourselves to a one-dimensional potential. The
anharmonic contributions are represented by the coefficients $b_1$ and $b_2$.
Whereas $b_1$ corresponds to the local anharmonicity around the origin $x=0$,
$b_2$ reflects the overall anharmonicity of the well. We therefore assume that
close to $x_c$ the term proportional to $b_2$ is much more relevant than the
term proportional to $b_1$. With simple algebra the anharmonic corrections to
the harmonic partition function as well as the specific heat of $V(x)$ can be
calculated in the limit of low and high temperatures. We obtain for low
temperatures
\be
\label{z_low}
z^{anh}(T) = 1 + \frac{3b_1 T}{2 a^2},
\ee
\be
\label{c_low}
c^{anh}(T) = \frac{3b_1T}{2a^2}
\ee
and for the limit $T \rightarrow \infty$ to lowest order in $1/T$
\be
\label{z_high}
z^{anh}(T) = \sqrt{\frac{12}{\pi}}\sqrt{\frac{V_c}{T}},
\ee
\be
\label{c_high}
c^{anh}(T) = -1
\ee
where $c^{anh}(T) = c(T) - c^{harm}(T)$. Here we defined
\be
\label{vc}
V_c \equiv V(\pm x_c)\approx (1/3) a^{1.5}b_2^{-0.5}
\ee
which corresponds to
the energy difference between maximum, corresponding to a saddle in the PEL, and
minimum.

It is straightforward to explain the temperature dependence of the specific
heat. From Eq.\ref{c_low} it is evident that there exist positive  anharmonic
contributions and ambient temperatures for which the anharmonicity is dominated
by the local anharmonicity term proportional to $b_1$. For some temperature
$T_{c,r}$, however, the system realizes the finite size of the potential well
and correspondingly the presence of an upper energy cutoff. This results in a
strong decrease of the specific heat until for very high temperatures the ideal
gas limit is recovered, i.e. vanishing configurational contribution to $c(T)$.
This effect is governed by the global anharmonicity term proportional to $b_2$.
Interestingly, $T_{c,r}$ is close to the temperature for which upon cooling the
PEL starts to become relevant (\cite{Sastry98,Heuer99} and Fig.7).

For explaining the anharmonicity effects related to the temperature dependence
of $G_{eff}(\epsilon)$ and $\langle \ln y^{harm} \rangle (\epsilon)$ additional
properties of the PEL have to be postulated: (i) The local anharmonicity, i.e.
$b_1$, only mildly depends on energy. This assumption is compatible with the
observation that also the local force constants, i.e. $a$, only show a very weak
dependence on energy; see Fig.10. (ii) Low-energy inherent structures possess
larger barrier heights, corresponding to larger values of $V_c$ in our simple
model potential. Evidence for this assumption have been presented in
\cite{Sastry98,Heuer99}.

First we deal with the apparent temperature dependence of the effective density
of inherent structures $G_{eff}(\epsilon)$. For the three lower temperatures we
already learned from analysis of the specific heat that local anharmonicity
effects are already present. According to assumption (i) the anharmonic
contribution only mildly depends on energy $\epsilon$. Therefore to a good
approximation these anharmonic effects are not visible in Fig.9 since they are
irrelevant for the scaling analyis. As discussed above only a strong
$\epsilon$-dependence of $z^{anh}(\epsilon,T)$ renders $G_{eff}$ temperature
dependend. For the two high temperatures however, where according to the
specific heat analysis the high-temperature expansion, i.e. Eq.\ref{z_high},
becomes relevant, the anharmonicity depends on $V_c$. Following assumption (ii)
the anharmonic contributions are significantly larger for low-energy inherent
structures. This leads to an overestimation of $G_{eff}(\epsilon)$ in the region
of low energies. This explains why the effective densities, obtained for
different temperatures by the above analyis, do not overlap at high
temperatures.

For elucidating the temperature dependence of $\langle \ln y^{harm} \rangle
(\epsilon)$ one has to take into account the variation of $Y_i^{harm}$ for
inherent structures with the same energy $\epsilon_i = \epsilon$.  According to
Eq.\ref{vc} one can expect that inherent structures with larger force constants
$a$, i.e. smaller $Y_i^{harm}$ possess somewhat larger barrier heights, i.e.
larger $V_c$. According to Eq.\ref{z_high} this results in frequent sampling of
inherent structures with small $Y_i^{harm}$. As a consequence the average value
$\langle
\ln y^{harm} \rangle (\epsilon)$ at fixed $\epsilon$ should decrease with
temperature at sufficiently high temperatures in agreement with the numerical
findings in Fig.10.  In summary, our simple model potential qualitatively
reproduces all anharmonicity features observed in our simulations.

{\bf Kauzmann temperature and finite-size effects}

The Kauzmann temperature $T_K$ has been introduced as the temperature for which
the configurational entropy of the glass-forming system would disappear in
equilibrium conditions. Thus knowledge of $G(\epsilon)$ enables one to estimate
$T_K$. For $T=T_K$ one expects the relaxation time to diverge since only a
single configuration is accessible. In analogy to phase transitions one might
expect modifications for finite systems: the Kauzmann temperature is smeared out
and for $T < T_K$ the system still has  a finite relaxation time.

In our case $G(\epsilon)$ is mainly determined by the parameters $\alpha,
\sigma$, and $N$. For $N=20$ the dynamics at low temperatures is also determined
by a single inherent structure. In what follows we restrict ourselves to a
perfect Gaussian distribution and consider the effects which arise from the fact
that at sufficiently low temperatures the system is sensitive to the fact that
one has a low-energy cutoff of $G(\epsilon)$, i.e. $G(\epsilon) = 0$ for
$\epsilon <
\epsilon_{min}$ one has $G(\epsilon)
= 0$ due to the finite (albeit exponential large) number of inherent structures.
A good indicator is the variance of $P(\epsilon,T)$. For large temperatures (but
not too large in order to avoid anharmonic effects, see above) one expects this
variance to be constant and identical to the variance of $G(\epsilon)$. In
contrast, for $T
\rightarrow 0$ the system is stuck in the inherent structure with the lowest energy, giving rise to
a vanishing variance. The temperature where this crossover occurs and which can
be identified as the Kauzmann temperature $T_K$ can be estimated by the
condition that the energy interval
$[\langle{\epsilon}\rangle_{T}-a\sigma,\langle{\epsilon}\rangle_{T}+a\sigma]$
($\langle{\epsilon}\rangle_T$: maximum of $P(\epsilon,T)$), for which the
distribution $P(\epsilon,T)$ has its main contributions, starts to approach the
value of $\epsilon_{min}$, i.e.
\be
\langle{\epsilon}\rangle_{T_K}-a\sigma = \epsilon_{min}.
\ee
$a$ is a constant of order unity. The strength of the dependence of $T_K$ on
this parameter $a$ is a meausure for the temperature width of the transition.
Thus one would expect that for large systems the dependence on $a$ vanishes; see
above.
 The value of $\epsilon_{min}$ is determined by
the condition $G(\epsilon_{min}) =1$. For a gaussian distribution the value of
$\langle \epsilon \rangle_T$ is given by
\be
\langle{\epsilon}\rangle_{T} = \langle{\epsilon}\rangle_{T=\infty} - \frac{\sigma^2}{T}.
\ee
Thus we obtain
\be
 \frac{1}{2\pi\sigma^2} \exp(-(-\sigma^2/T_K - a\sigma)^2/2\sigma^2) \exp{\alpha N} = 1
\ee
Neglecting corrections of order $1/N$ this relation can be rewritten as
\be
\label{det_tk}
\frac{\sigma / \sqrt{N}}{T_K} = \sqrt{2\alpha} - a/\sqrt{N}.
\ee
For large systems the last term disappears and thus $T_K$ is independent of the
value of $a$ in agreement with expectation. We do not know the value of $\alpha$
for systems larger than $N =30$. However, since already for $N
\ge 60$ the parameter $\sigma^2/N$ (Fig.11) and ${\epsilon}_{max}/N$ (Fig.7) have reached
their limiting value one may speculate that together with the values of $\alpha$
for $N=20$ and $N=30$ the  value of $\alpha$ for large $N$ is larger than 0.7
and smaller than 1.2 (linear extrapolation). On this basis the Kauzmann
temperature can be estimated as $T_K = 0.39 \pm 0.05$. As a comparison  the
mode-coupling critical temperature  has been estimated for the present system as
$T_c = 0.56$; see Ref.\cite{Kob95}, taking into account the temperature shift of
30\% (see below). For smaller systems the additional term $a/\sqrt{N}$ clearly
incraeses the value of $T_K$. As has been already discussed in the context of
Fig.11 the dynamics at the three lower temperatures for $N=20$ is already
significantly influenced by the presence of the lower cutoff of $G(\epsilon)$. A
quantitative analysis, however, is hampered by the fact that the structure of
$G(\epsilon)$ close to the lower cutoff is more complicated due to the presence
of a single or a few inherent structure, dominating the physics; see also
Ref.\cite{Stillinger88}. Summarizing this line of argumentation, the
N-dependence of $T_K$ as expressed in Eq.\ref{det_tk} clearly leads to finite
size effects and it is exactly this type of finite size effect which we have
explicitly found in our simulations. Finally we note that this derivation is
similar to what has been done for the random energy model \cite{Derrida80}.

Very recently, Kim and Yamamoto have analysed soft sphere systems and found a
significant finite size effect when comparing systems with $N=108$ and $N=10000$
particles \cite{Kim99}. The interaction of adjacent particles in LJ systems
under high pressure is dominated by the first term proportional to $r^{-12}$.
Therefore it is reasonable to assume that the physics of very dense LJ systems
is somewhat similar to that of soft sphere systems. Recent work on monatomic
LJ-type systems \cite{Heuer97f} as well as theoretical predictions
\cite{Stillinger99} show that the number of inherent structures strongly
decreases with increasing pressure. In our terminology this would result in a
much smaller value of $\alpha$ for LJ systems at high density and thus soft
sphere systems than for LJ systems at ambient densities, discussed in this work.
According to the above discussion of Eq.\ref{det_tk} this would mean that finite
size effects, related to the finite range of energies of inherent structures,
occur for much larger $N$ as compared to LJ-type glasses. In contrast, Kim and
Yamamoto have explained their finite size effect on the basis of dynamic
heterogeneities, i.e. the presence of fast and slow particles. Finite size
effects were observed at a temperature for which the length scale $\xi$ of
dynamic heterogeneities, i.e. the cluster size of slow or fast particles, became
as large as the simulation box. The interesting question arises whether the
temperature for which $\xi$ is of the order of the box size is strongly related
to the temperature for which the finite number of inherent structures, i.e. the
energy $\epsilon_{min}$ becomes relevant. This picture would be consistent with
the notion that for macroscopic systems the length scale of the glass transition
diverges at the Kauzmann temperature.

{\bf Physical picture}

Based on our results as well as previous work on PELs the following picture
seems to emerge. Coming from low temperatures the system mainly stays close to
the inherent structures and the dynamics can be described by a superposition of
local vibrations and hopping processes. Around a temperature close to the
mode-coupling temperature $T_c$ {\it local} anharmonic effects start to play a
role as seen, e.g., from the temperature dependence of the
mean-square-displacement around one inherent structure \cite{Sastry98}, from the
comparison of the inherent and the real trajectories \cite{Schroder99}, and from
the presence of anharmonic contributions of the specific heat above $T_c$, seen
in this work. Despite the anharmonic effects, the PEL still has a strong
influence on the dynamics as explicitly shown in Ref.\cite{Buechner99}. At a
temperature of the order $ 2 T_c$ {\it global} anharmonic effects start to
dominate the dynamics which are partly related to the presence of saddles
between inherent structures and thus to the finite size of the basins of
attraction. It is, of course, still the PEL, representing the total potential
energy of the system, which is responsible for the dynamics. However, the
topography of the individual inherent structures, including their close
neighborhood, becomes irrelevant \cite{Buechner99}.

In summary, we have obtained a thermodynamic picture of LJ-type glasses  based
on an appropriate numerical analysis of the PEL. Questions concerning the
Kauzmann temperature, finite-size effects, and anharmonicities have been
approached. The present work is a step in elucidating the nature of the
supercooled state on the basis of the PEL, which hopefully stimulates further
research along this direction.

We gratefully acknowledge helpful discussions with B. Doliwa, H.W. Spiess, and
K. Binder. After finishing this work we learned about simultaneous independent
activities by F. Sciortino, W. Kob, and P. Tartaglia along a similar line of
thought \cite{Sciortino99}. This work was supported by the DFG via the SFB 262.

\setcounter{figure}{0}
\begin{figure}
\epsfxsize=5in
\epsfbox{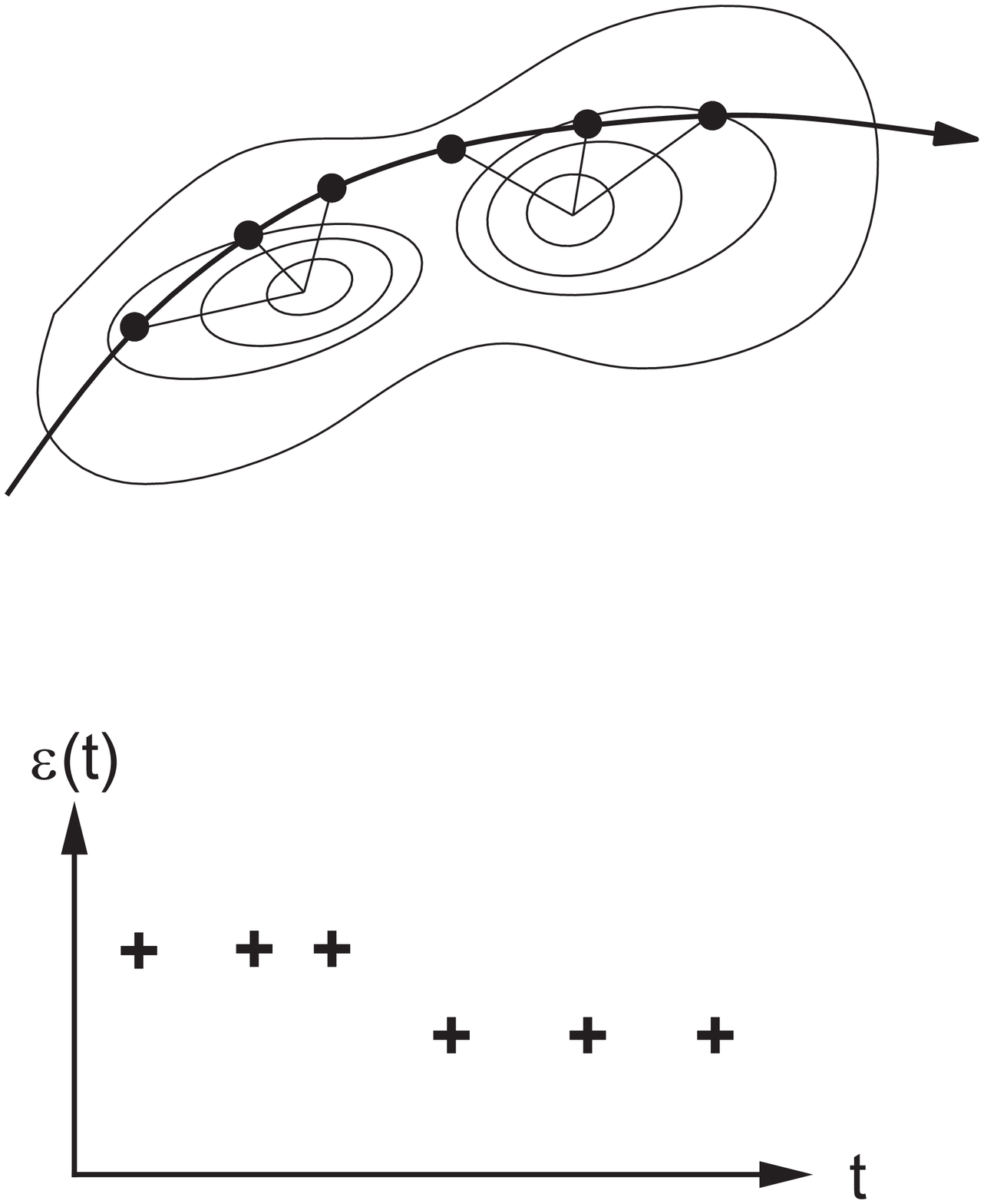}
\vspace{2cm}
\caption{
Schematic presentation of the algorithm. On a regular basis MD configurations
are quenched, giving information about the energy $\epsilon(t)$ of the
corresponding inherent structure. }
\label{fig:fig1}
\end{figure}

\newpage

\begin{figure}
\epsfxsize=5in
\epsfbox{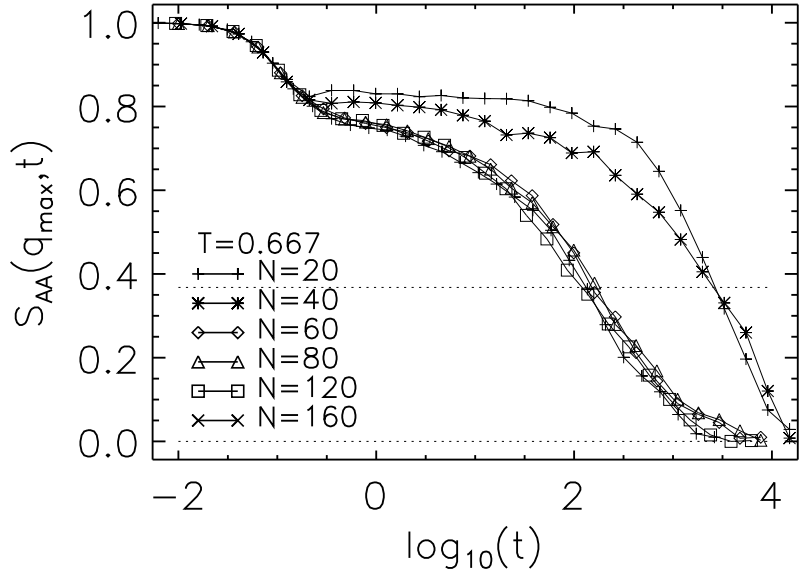}
\vspace{2cm}
\caption{
The incoherent scattering function $S_{AA}(q,t)$ for $T=0.667$ for
different system sizes $N$, ranging from $N=20$ to $N=160$.
}
\label{fig:fig2}
\end{figure}

\newpage

\begin{figure}
\epsfxsize=5in
\epsfbox{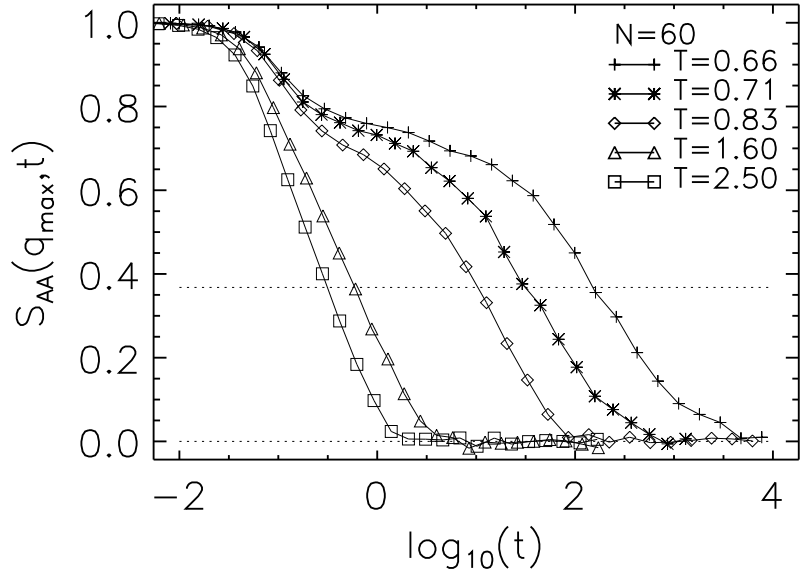}
\vspace{2cm}
\caption{
The temperature dependence of the incoherent scattering function $S_{AA}(q,t)$ for $N=60$. }
\label{fig:fig3}
\end{figure}

\newpage

\begin{figure}
\epsfxsize=5in
\epsfbox{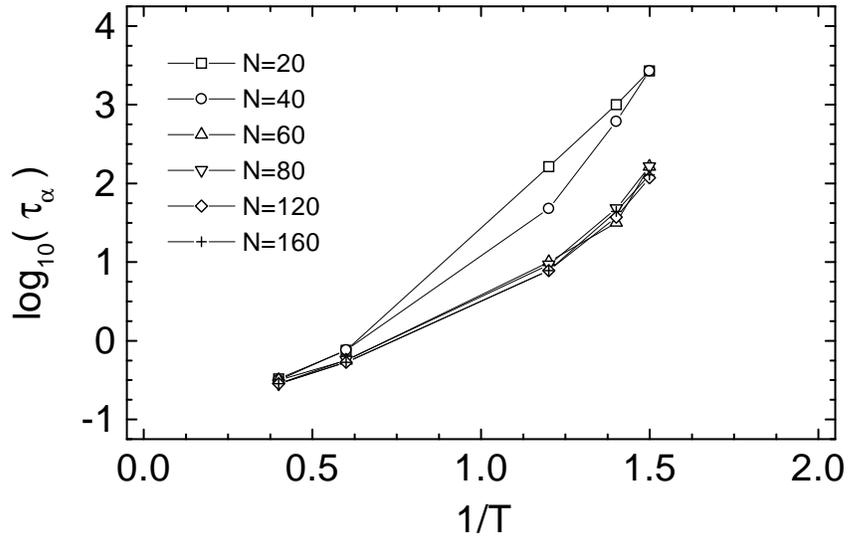}
\vspace{2cm}
\caption{
The $\alpha$-relaxation time for different temperatures and system sizes,
determined by the condition $S(q,\tau_\alpha) = 1/e$. }
\label{fig:fig4}
\end{figure}

\newpage

\begin{figure}
\epsfxsize=5in
\epsfbox{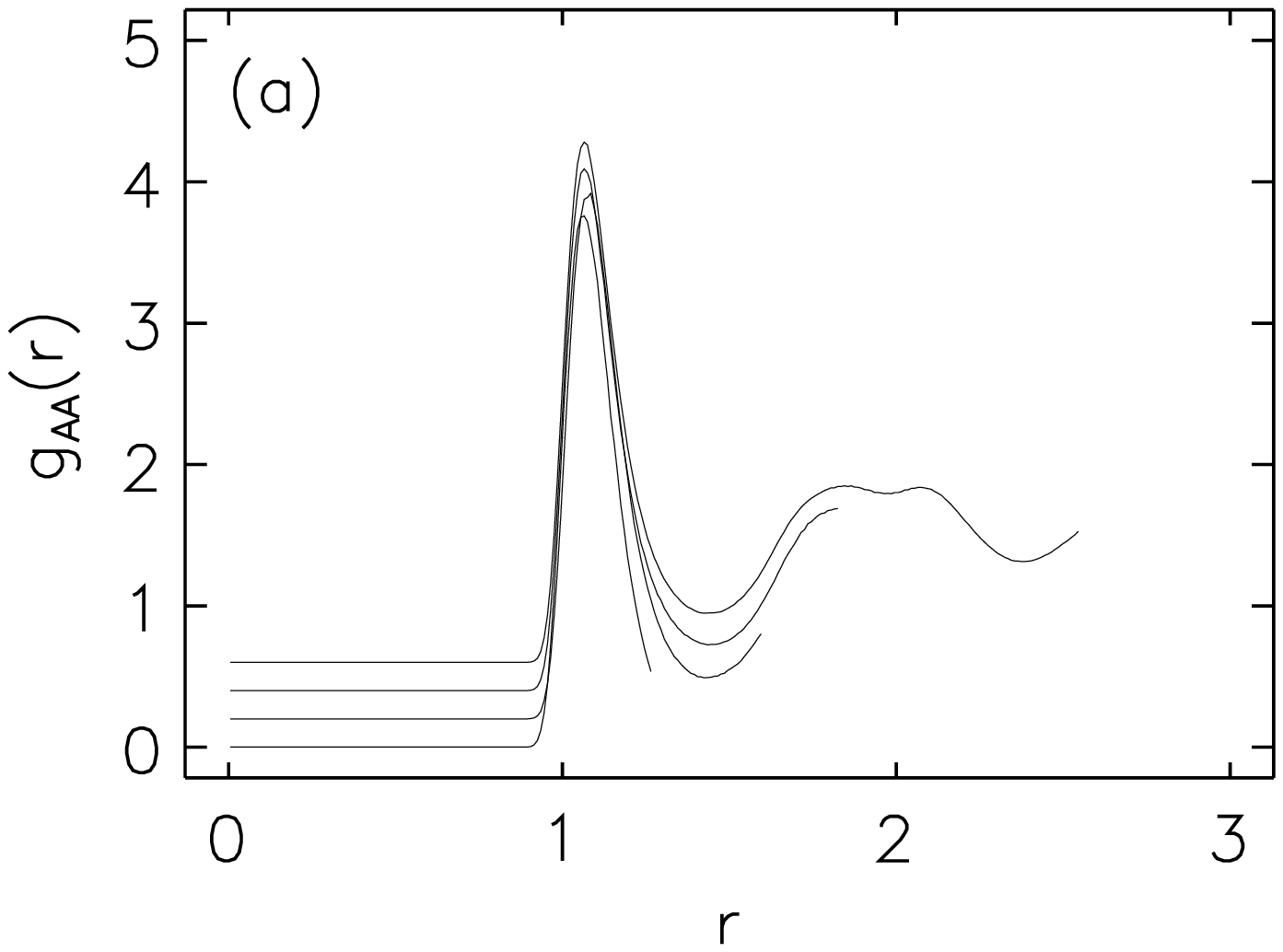}
\end{figure}
\begin{figure}
\epsfxsize=5in
\epsfbox{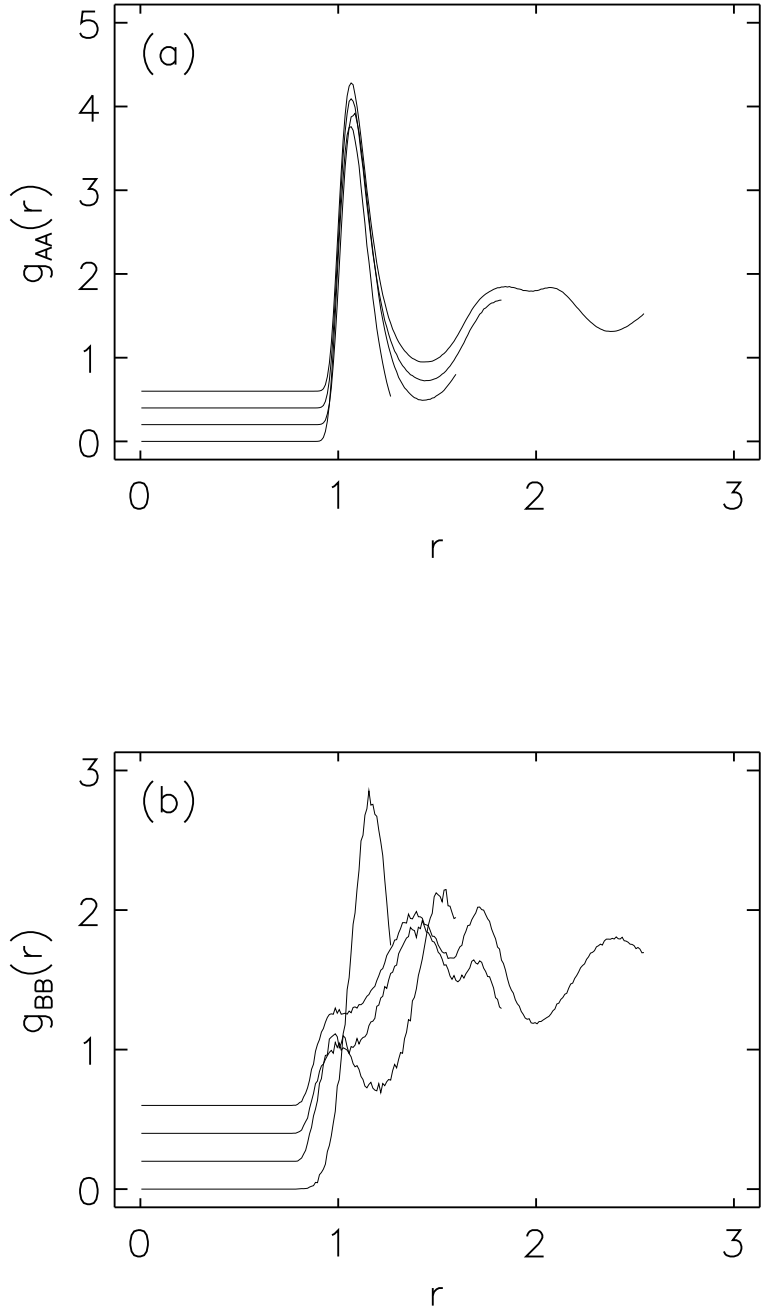}
\vspace{2cm}
\caption{
The pair correlation functions (a) $g_{AA}(r)$ and (b) $g_{BB}(r)$ for  system
sizes $N=20,40,60,160$, determined for $T = 0.667$. The offset has been shifted
for better comparison. }
\label{fig:fig5}
\end{figure}

\newpage

\begin{figure}
\epsfxsize=3in
\epsfbox{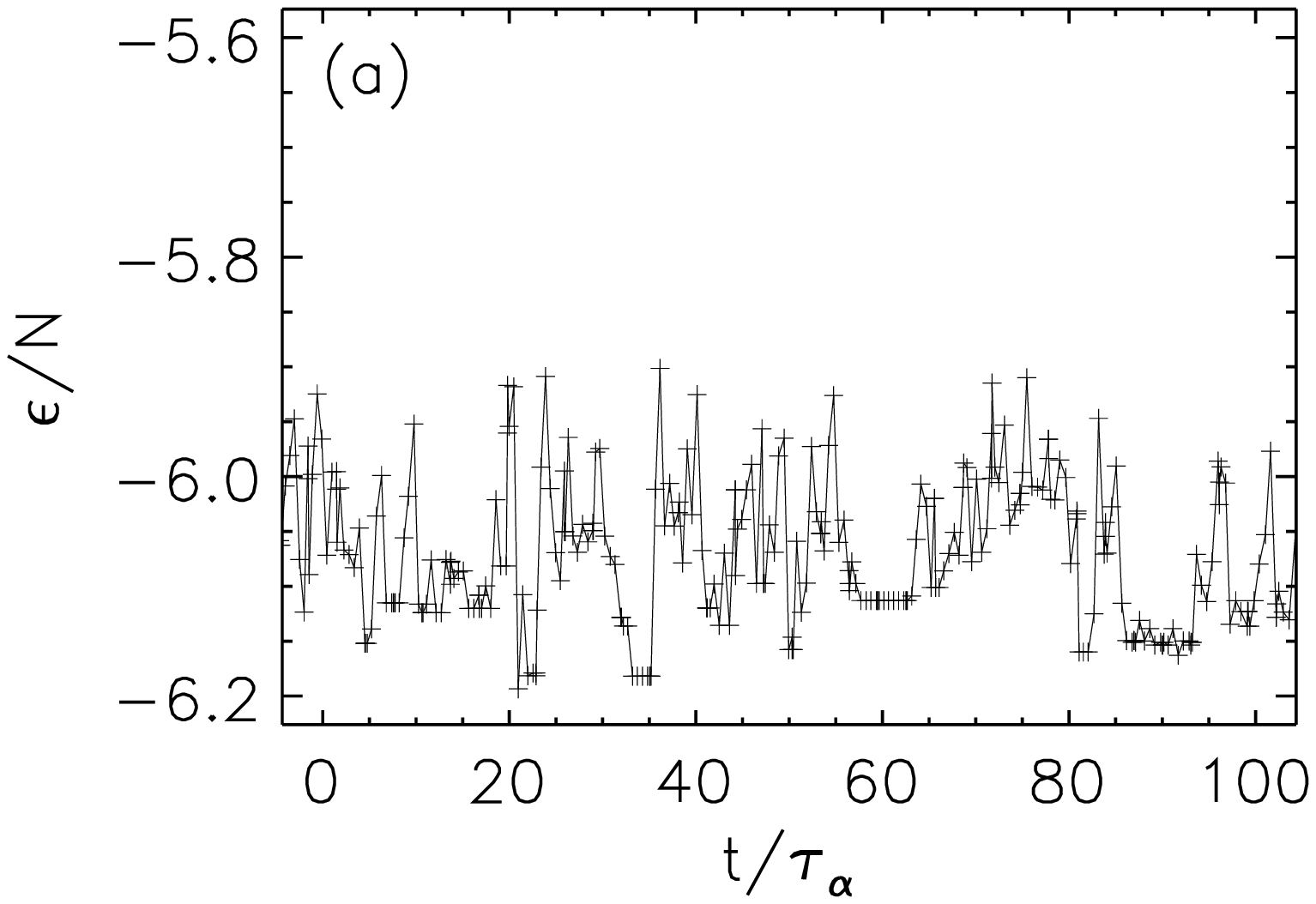}
\end{figure}
\begin{figure}
\epsfxsize=3in
\epsfbox{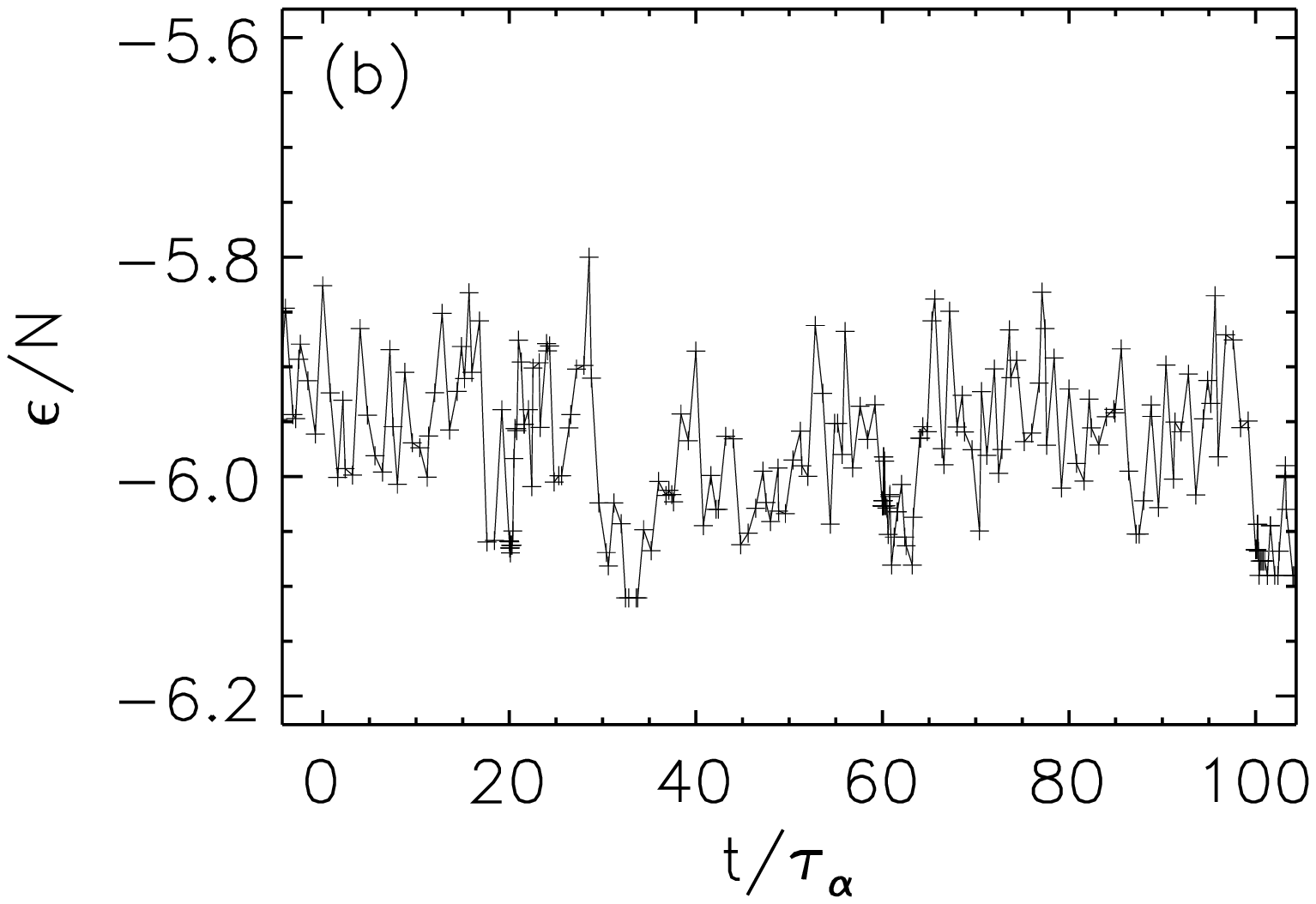}
\end{figure}
\begin{figure}
\epsfxsize=3in
\epsfbox{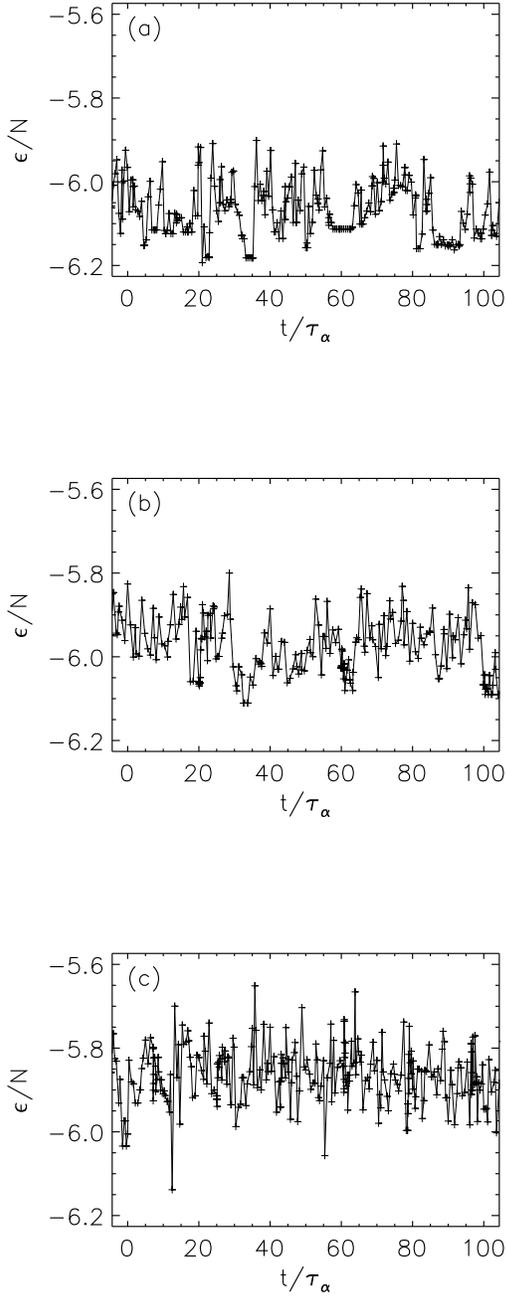}
\vspace{2cm}
\caption{
The time-dependence of the energy of inherent structures
$\epsilon(t)$ for three representative temperatures (a)$T=0.667$, (b) $T=
0.833$, (c) $T= 1.667$ and
 for system size $N=60$.
}
\label{fig:fig6}
\end{figure}

\newpage

\begin{figure}
\epsfxsize=5in
\epsfbox{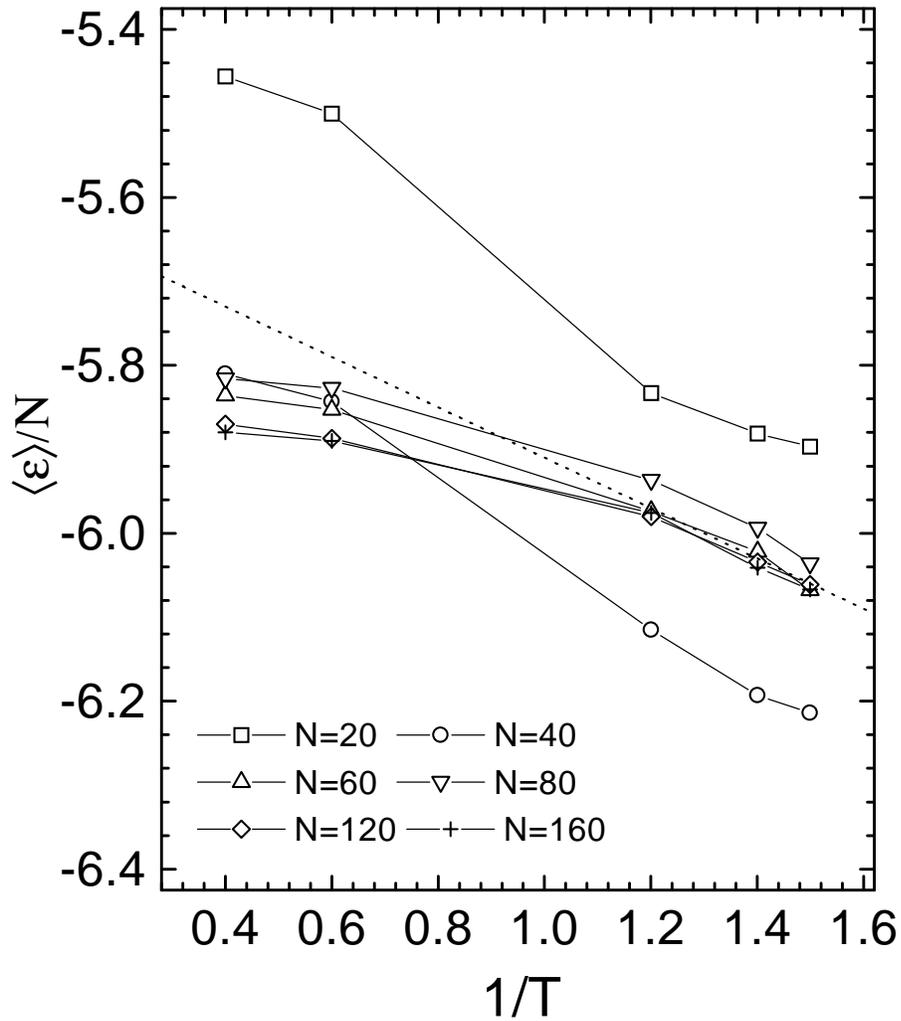}
\vspace{2cm}
\caption{
The average value of the energy of inherent structures $\langle
\epsilon \rangle_T$ for different temperatures and different system sizes. The
solid line corresponds to an estimation for $N=60$, based on
$G_{eff}(\epsilon)$, see Fig.9. }
\label{fig:fig7}
\end{figure}

\newpage

\begin{figure}
\epsfxsize=5in
\epsfbox{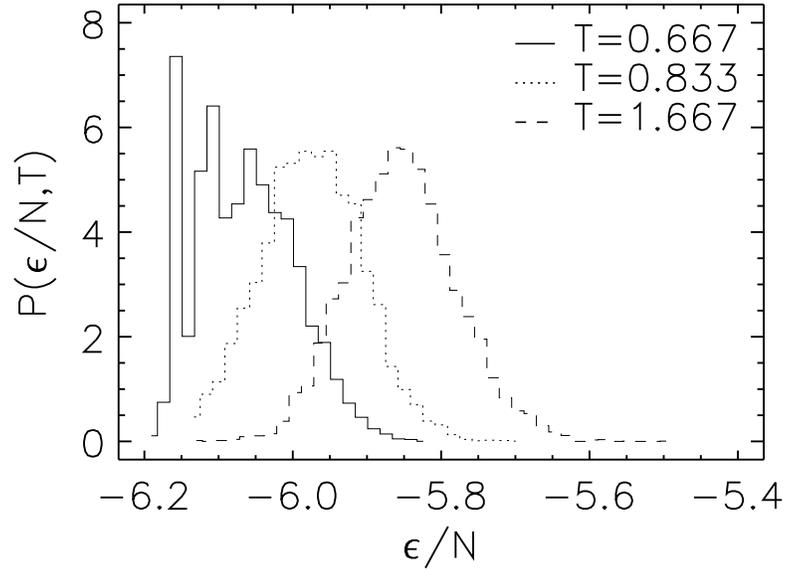}
\vspace{2cm}
\caption{
The distribution $P(\epsilon,T)$ of inherent structures at three
different temperatures ($T=0.667, 0.833, 1.667$ from left to right).
}
\label{fig:fig8}
\end{figure}

\newpage

\begin{figure}
\epsfxsize=5in
\epsfbox{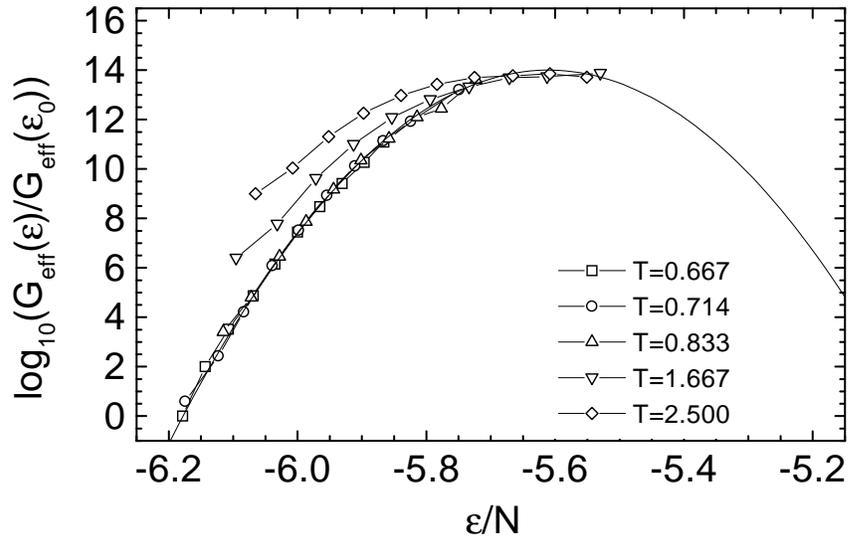}
\vspace{2cm}
\caption{
Determination of $G_{eff}(\epsilon)$ on the basis of $P(\epsilon,T)$ for $N=60$.
The individual curves have been shifted in order to obtain an optimum overlap. }
\label{fig:fig9}
\end{figure}

\newpage

\begin{figure}
\epsfxsize=5in
\epsfbox{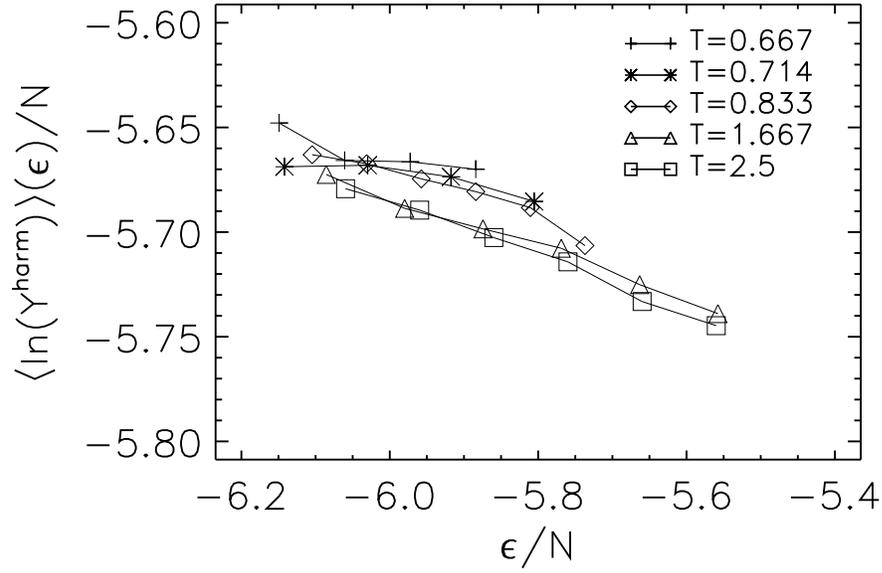}
\vspace{2cm}
\caption{
The average $\langle \ln y^{harm} \rangle(\epsilon)$ evaluated at different
temperatures in dependence on energy. Note that small values of $y^{harm}$ correspond
to large force constants around the respective inherent structures.}
\label{fig:fig10}
\end{figure}

\newpage

\begin{figure}
\epsfxsize=5in
\epsfbox{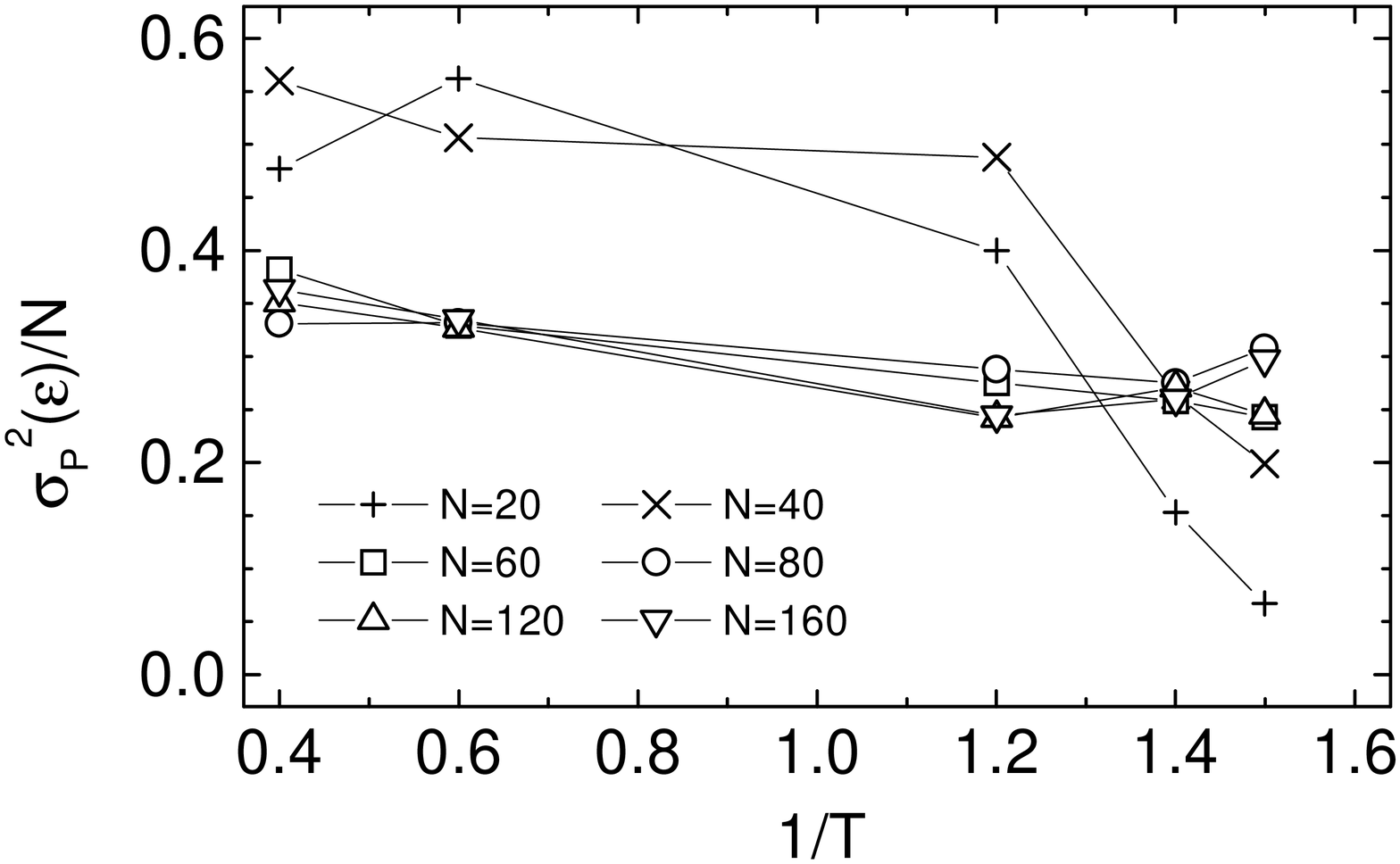}
\vspace{2cm}
\caption{
The variance $\sigma_P^2(T)$ of $P(\epsilon,T)$ calculated for different
temperatures and system sizes. The strong temperature dependence for $N=20$ and $N=40$ is 
explained in the text.}
\label{fig:fig11}
\end{figure}

\newpage

\begin{figure}
\epsfxsize=3in
\epsfbox{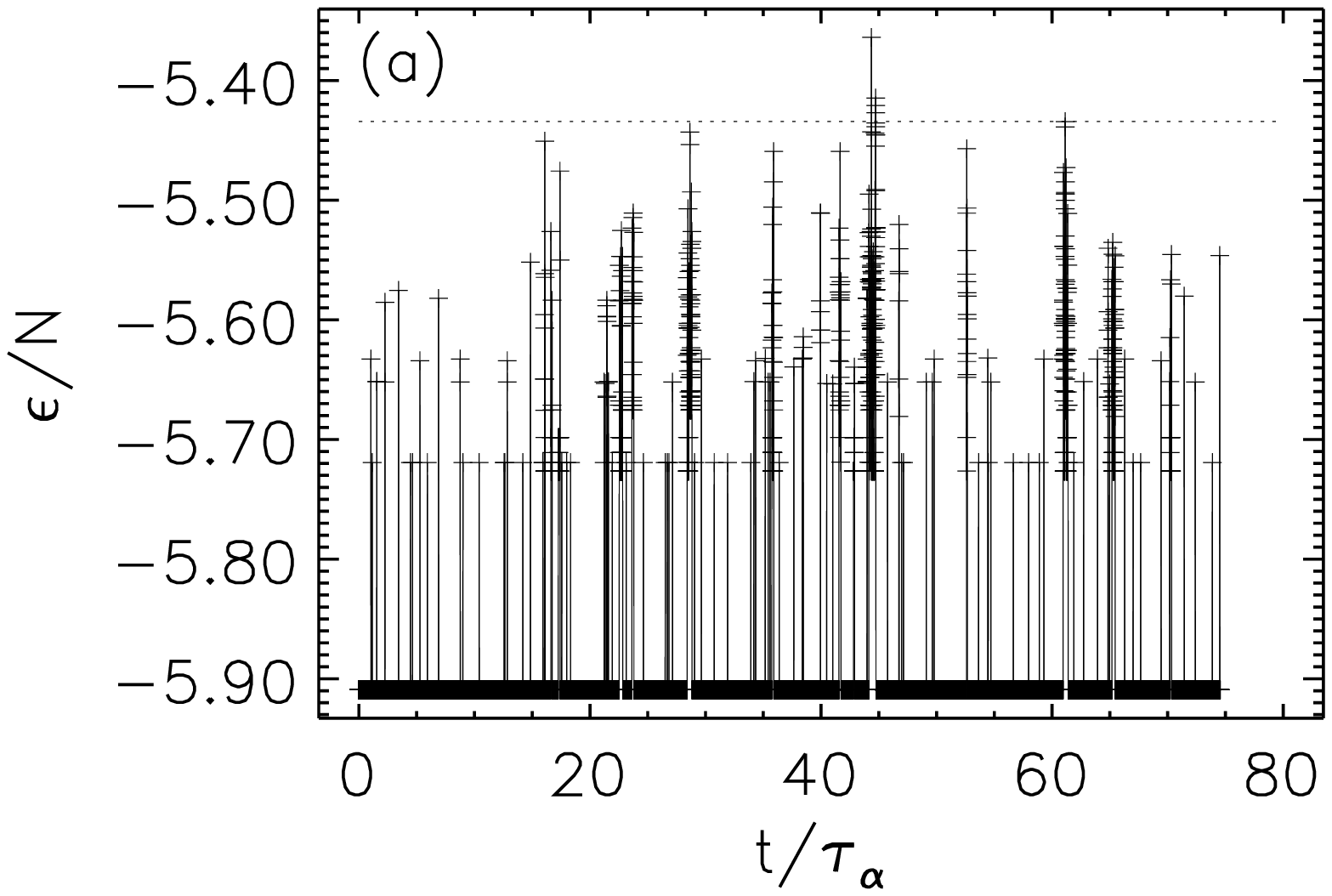}
\end{figure}
\begin{figure}
\epsfxsize=3in
\epsfbox{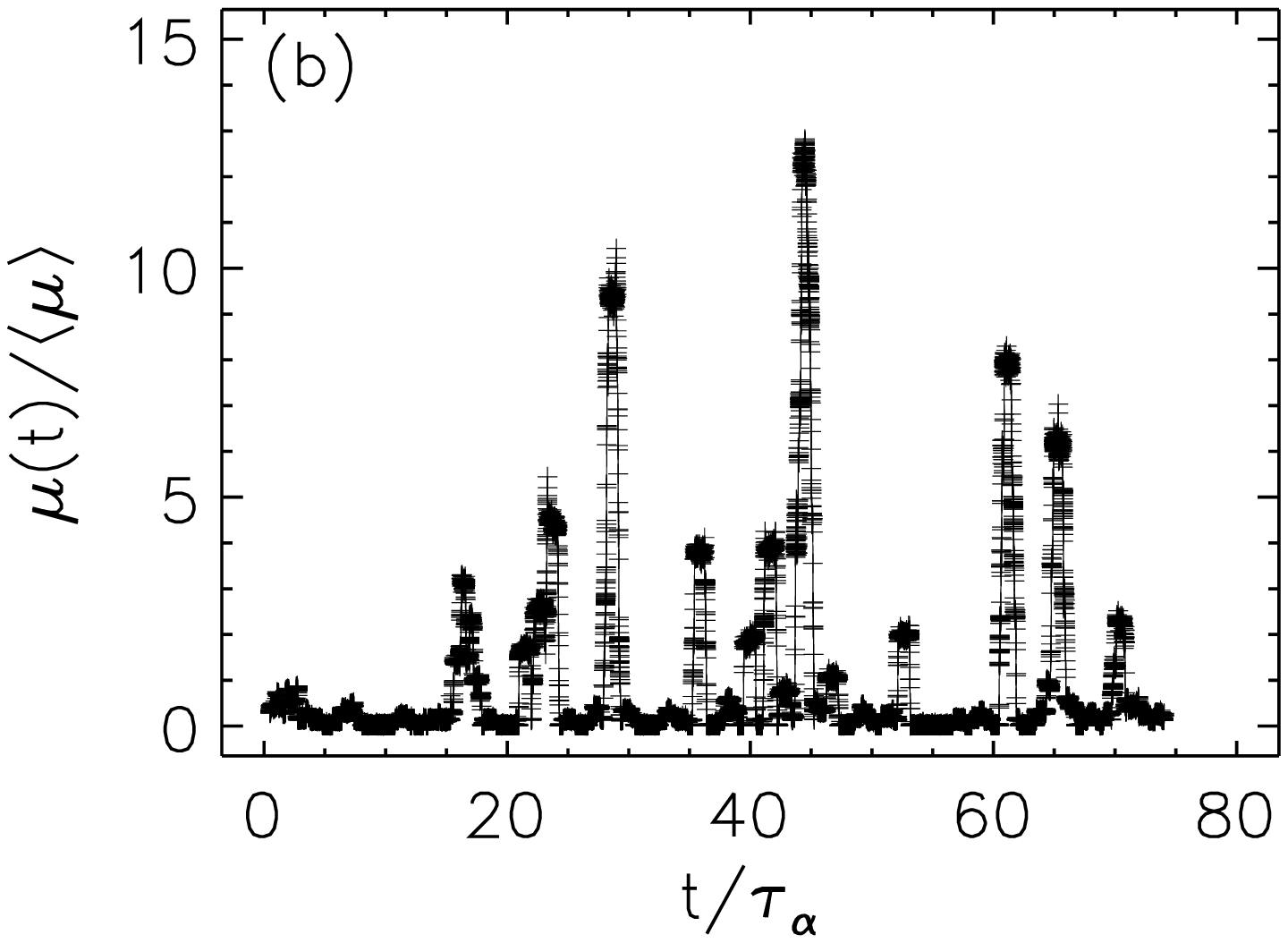}
\end{figure}
\begin{figure}
\epsfxsize=3in
\epsfbox{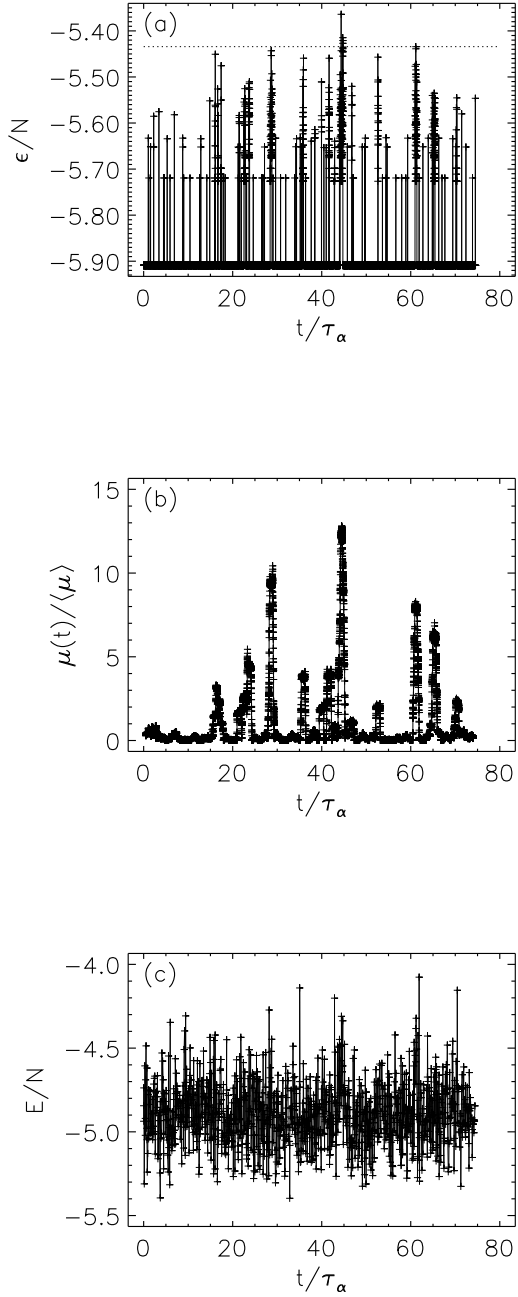}
\vspace{2cm}
\caption{
(a) The time series of the energy of inherent structures $\epsilon(t)$ for
$N=20$ at $T=0.833$; the broken line indicates the activation energy of the
dynamics at low temperatures; see Fig.4; (b) the corresponding time series of
mobilities $\mu(t)$; (c) the corresponding time series of the energy of the MD
configurations $E(t)$. }
\label{fig:fig12}
\end{figure}

\newpage

\begin{figure}
\epsfxsize=5in
\epsfbox{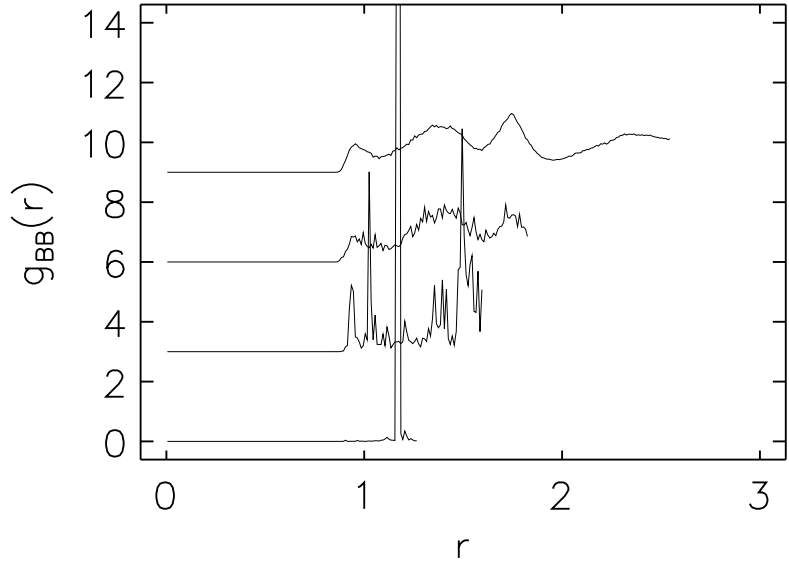}
\vspace{2cm}
\caption{
The pair correlation function $g_{BB}(r)$ for $N=20,40,60,$ and $160$ at $T=0.833$ determined
from the inherent structures. }
\label{fig:fig13}
\end{figure}

\newpage

\begin{figure}
\epsfxsize=5in
\epsfbox{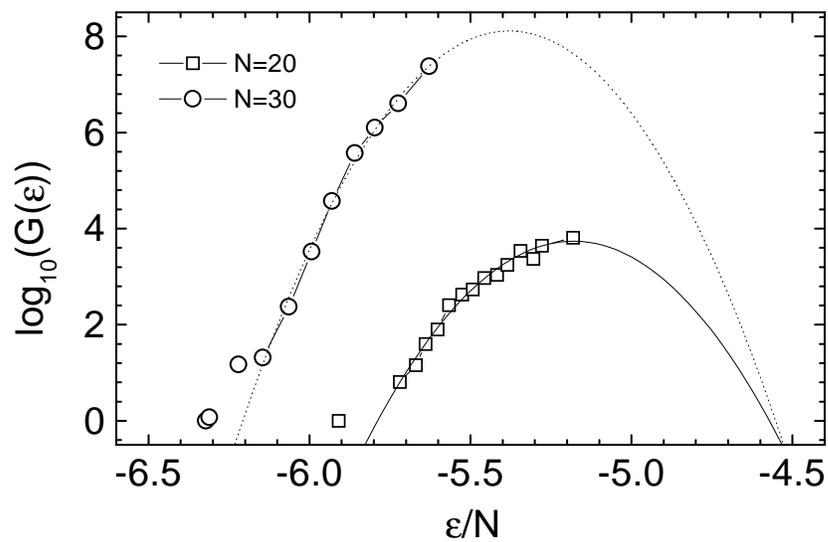}
\vspace{2cm}
\caption{
The density of inherent structures $G(\epsilon)$ for $N=20$ and $N=30$ obtained from 
simulations at a single temperature.  }
\label{fig:fig14}
\end{figure}

\newpage

\begin{figure}
\epsfxsize=5in
\epsfbox{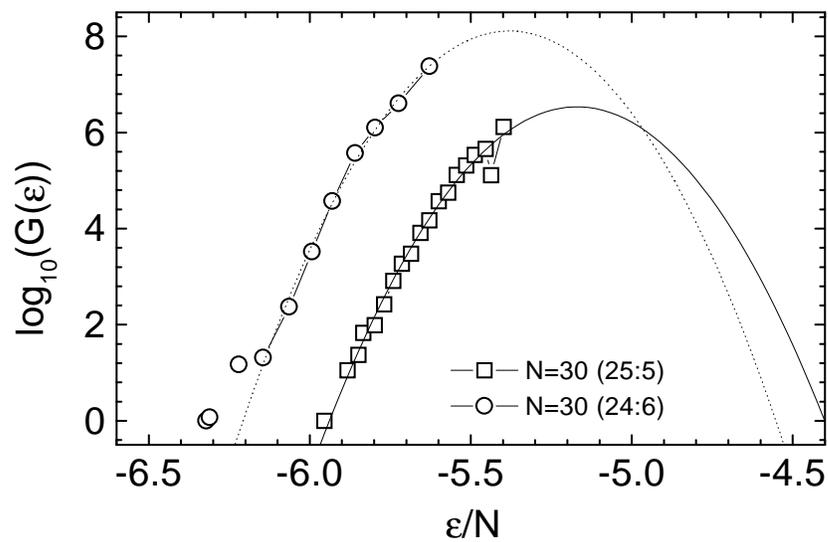}
\vspace{2cm}
\caption{
 The density of inherent structures $G(\epsilon)$ for two different
compositions ($N_A=25,N_B=5$ vs. $N_A=24,N_B=6$). }
\label{fig:fig15}
\end{figure}

\newpage

\begin{figure}
\epsfxsize=5in
\epsfbox{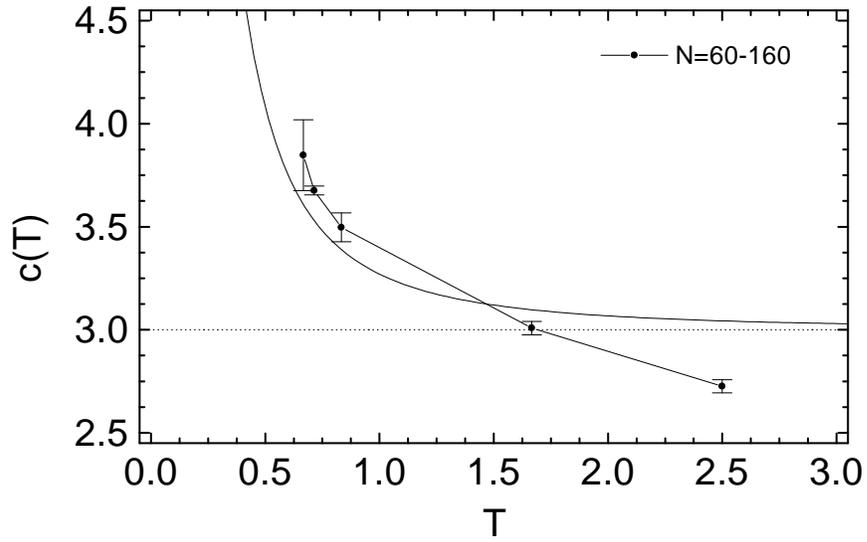}
\vspace{2cm}
\caption{
The specific heat as obtained from $G_{eff}(\epsilon)$ and averaged over all
system sizes $N \ge 60$ together with the actual specific heat obtained from
analysis of the energy fluctuations in the MD simulation. The deviations
correspond to anharmonic contributions. }
\label{fig:fig16}
\end{figure}

\newpage

\begin{figure}
\epsfxsize=5in
\epsfbox{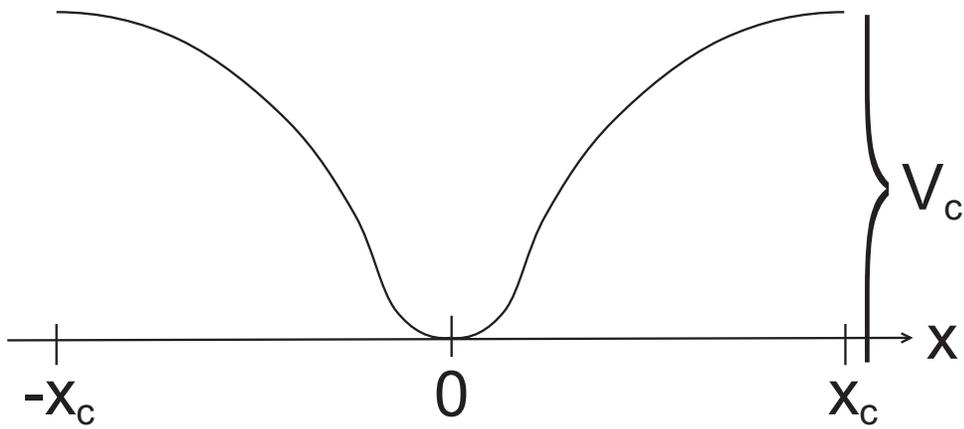}
\vspace{2cm}
\caption{
Sketch of the model potential V(x) as described in the text. The size of the basin of attraction
and the potential height are indicated. }
\label{fig:fig17}
\end{figure}

\end{document}